\documentclass[hyperref,twocolumn]{IEEEtran}
\usepackage[english]{babel}
\usepackage[utf8]{inputenc}
\usepackage[T1]{fontenc}
\usepackage{array}
\usepackage{graphicx}
\usepackage{color}
\usepackage{amsmath}
\usepackage{cases}
\usepackage{amssymb}
\usepackage{algorithm}
\usepackage{algpseudocode}
\usepackage{hyperref}

\newcommand{\RR}{\mathbb{R}} 

\newcommand{\NN}{\mathbb{N}}
\newcommand{\bd}{\mathbf}

\DeclareMathOperator*{\argmax}{arg\,max}

\newcommand{\change}[1]{{#1}}

\definecolor{darkviolet}{rgb}{0.58,0,0.83} 

\renewcommand{\paragraph}[1]{\textbf{#1.}}

\begin{document}

\title{\change{Inpainting of long audio segments with similarity graphs}}
\author{\IEEEauthorblockN{Nathana\"el Perraudin$^*$\thanks{$^*$N. Perraudin and N.Holighaus contributed equally to this work.}\IEEEauthorrefmark{2}, Nicki Holighaus$^*$\IEEEauthorrefmark{3}, Piotr Majdak\IEEEauthorrefmark{3} and Peter Balazs\IEEEauthorrefmark{3}}\\
\IEEEauthorblockA{\IEEEauthorrefmark{2}Swiss Data Science Center, EPFL and ETH Zürich, Switzerland}\\
\IEEEauthorblockA{\IEEEauthorrefmark{3}Acoustics Research Institute, Austrian Academy of Sciences, Wohllebengasse 12–14, 1040 Vienna, Austria}
}
\maketitle


\begin{abstract}
We present a novel method for the compensation of long duration data \change{loss} in 
audio signals, in particular music. The concealment of such signal defects is
based on a graph that encodes signal structure in terms of time-persistent 
spectral similarity. A suitable candidate segment for the substitution of
the lost content is proposed by an intuitive optimization scheme
and smoothly inserted into the \change{\emph{gap}, i.e. the lost or distorted signal region.}
Extensive listening tests show that the 
proposed algorithm provides highly promising results when applied to a variety
of real-world music signals.
\end{abstract}

\section{Introduction}
\change{The loss or corruption of entire segments of audio data
is a highly important problem in music enhancement and restoration. 
Such corruptions can range from short bursts in the range of few 
milliseconds to extended distortions that persist over several hundred
or even thousands of milliseconds. Short distortions such as clicks
or clipping have seen extensive coverage in the literature~\cite{siedenburg2013audio,adler2011constrained,godsill2002digital}, while
the concealment of moderate length distortions, roughly in the range of 
$10$ to at most $100$~ms, is treated in packet loss compensation~\cite{perkins1998survey,bahat2015self}
and previous work on audio inpainting~\cite{adler2012audio,siedenburg2013audio,lagrange2005long}. For such corruptions, it 
is often reasonable to assume that the lost signal is almost stationary
for the duration of the corruption and/or can be inferred from the reliable
information surrounding the unreliable segment. For longer duration loss, 
such an assumption is increasingly unrealistic and a restoration technique
cannot rely only on local information. Here, we propose a method to
compensate for such extended data loss by considering information from the
entirety of uncorrupted audio available.

Data loss or corruption in the range of seconds can have various causes,
e.g. partially damaged physical media, such as phonograph cylinders, 
shellac or vinyl records or even magnetic tapes. In live music recordings, 
imperfections due to unwanted noise sources originating from the audience, 
the artists themselves or the environment are quite common. Even in audio 
transmission, a short, but total, loss of the connection between transmitter 
and receiver may lead to data loss beyond just a few hundred milliseconds.
In each of these scenarios, the data loss has highly unpleasant consequences 
for a listener, and it is usually not feasible to reconstruct the lost content
from local information only. 

Previous work on concealment of data loss in audio, though mostly considering shorter 
corruption duration, has been performed under various names, depending on the target application 
and the employed methodology:
Audio inpainting~\cite{adler2012audio}, audio interpolation~\cite{Etter1996:Interpolation_AR}, waveform 
substitution~\cite{goodman1986waveform}, or imputation~\cite{smaragdis2009missing} to name but a few.
We will use the terminology of \emph{audio inpainting} in the remainder of this contribution. } 
When missing parts have a length no longer than $50$ms, sparsity-based techniques can be 
successful~\cite{adler2012audio,siedenburg2013audio,adler2011constrained}. 
Otherwise, techniques relying on auto-regressive 
modeling~\cite{Etter1996:Interpolation_AR}, sinusoidal 
modeling~\cite{lagrange2005long,lukin2008:parametric.interp.gaps} or based on 
self-content~\cite{bahat2015self} have been proposed. The latter provided promising results 
for speech signals with distortions up to $0.25$ seconds, while the former rely on a
simple signal model that does not comply with complex music signals. 

In this contribution, we propose a new algorithm, specifically
targeted at the concealment of long duration distortions in the range of several seconds 
\change{given a single piece of music. The task of determining distortion locations
is highly application-dependent and may be anything from trivial to very difficult. 
For the sake of focus, we assume the location of the distortion to be known}. 
Our method arises from the assumption that, across many musical genres,
the repetition, or variation, of distinct and recurring patterns (themes, melodies, rhythms, etc)
is a central stylistic element and thus heavily featured. When listening to music, 
we detect and memorize such \emph{internal redundancies}, thereby learning the mid- 
and large-scale structures of a music piece \cite{mcadams1987}. The exploitation of
such redundancies in the computational analysis and processing of music seems only 
natural and, indeed, has been proposed before, see e.g.\change{~\cite{foote1999visualizing,jehan2005creating,jehan2004event} 
or~\cite{rafii2013repeating}. The latter also provides a more extensive discussion of
repetition as an essential element of many musical genres.}
Although music information retrieval (MIR) provides many sophisticated 
methods for the analysis of micro- and macroscopic structures in music, properly handled, a simple \emph{time-frequency analysis} can provide 
all the necessary information to uncover significant similarities in music 
signals. 
The contributions of this work are the design of appropriate time-frequency features and their 
use for generating a map of similarities in music signals, as well as the use of the generated 
similarity map to drive the automatic concealment of long duration data loss.

\subsection{Related Work}
\change{Self-similarity in music has previously been employed in several areas of music analysis and processing, e.g. beat estimation and segmentation, and is often based on similarity matrices, as proposed by Foote~\cite{foote1999visualizing}. The similarity matrix can be constructed from various features, see e.g. \cite{foote2001beat,bartsch2001catch,cooper2002automatic,foote2000automatic}. Self-similarity has also been successfully used for music/voice separation and speech enhancement~\cite{rafii2013repeating,rafii2013online}. Finally, the automatic analysis of musical structure based on 
similarities is already found in~\cite{silva2016simple}, where it was used across songs for cover song detection. An alternate approach can be found in~\cite{jehan2005creating,jehan2004event}\footnote{These studies led to the 
founding of "The Echo Nest", see \url{http://the.echonest.com/}, 
a company specialized into audio feature design. The idea of a similarity 
graph already appears in the infinite jukebox: \url{http://labs.echonest.com/Uploader/index.html}.}. 
There, the division of music into short, rhythm-dependent pieces is proposed, 
each of which is supposed to correspond to a single \emph{beat}. Local features are obtained for 
each piece by combining previously established rhythm, timbre and pitch features, but the implementation details of their method are not disclosed. In this contribution, we propose a simple time-frequency feature built from the short-time Fourier magnitude and phase that implicitly encodes rhythmic, timbral and pitch characteristics of the analyzed signal all at once. We build a sparse similarity graph based on this feature that highlights only the strongest connections in a music piece. This similarity graph can be seen as a post-processed variant of Foote's 
similarity matrix and is used to perform data loss concealment by detecting suitable transitions between similar segments in a piece of music.}

The audio inpainting problem has mainly been addressed from a sparsity point of 
view. \change{The hypothesis is that audio is often approximately sparse in a time-frequency 
representation, i.e. it can be estimated using only a few 
time-frequency atoms.} Using classical $\ell_0$ or $\ell_1$ optimization 
techniques, algorithms have been designed to inpaint short audio 
gaps~\cite{adler2012audio,siedenburg2013audio}. Such methods strive for approximate
recovery of the lost data by sparse approximation in a time-frequency representation such 
as the short-time Fourier transform (STFT). Both their numerical and perceptual restoration quality
quickly degrade when the duration of lost data intervals exceeds $10$~ms. When applied to significantly longer gaps, these methods will simply fade out/in at the gap border and introduce silence in the inner gap region. 
Audio inpainting is known as "waveform substitution"~\cite{goodman1986waveform} by the community addressing packet loss recovery techniques~\cite{perkins1998survey}. Most packet loss methods, however, are naturally designed for low delay processing and compromise computation speed over quality, see also~\cite{bahat2015self} for a short overview. In that contribution, Bahat et al. propose an algorithm searching for similar parts of the signal using time-evolving features, conceptually resembling our own contribution. The method in~\cite{bahat2015self} 
is designed for packet loss concealment in speech transmission, however, and was tested only on gaps up to $0.25$ seconds. The reliance on Mel frequency cepstral coefficients (MFCC) is a good match for speech, but not optimally suited for music. In another approach, Martin et al.~\cite{martin2011exemplar} proposed an inpainting algorithm taking advantage of the redundancy in tonal feature sequences of a music piece. Their method is able to conceal defects with a length of several seconds, but performance of this algorithm depends on the amount of repetitive tonal sequences in a music piece \cite{martin2011exemplar} and it was only applied when a recurrence of the lost tonal sequence was present in the reliable signal. \change{It should be noted that parallel work on audio inpainting using self-similarity by Manilow and Pardo~\cite{manilow2017leveraging} has been presented while the present manuscript was under review.}

\subsection{Structure of the paper}
After the introduction, we introduce the idea of the similarity graph, Section \ref{sec:DNA}. The general method and construction of the graph is presented in Section \ref{sec:algo}. Technical details about the graph construction such as the exact choice of features and parameters are deferred to Section \ref{sec:TecDecGraph}. In Section \ref{sec:TecDecInp}, we detail how the similarity graph can be used for audio inpainting. Finally, the performance of the algorithm is discussed, based on both a basic verification experiment and though extensive listening tests, Section \ref{sec:evaluation}.

\section{A transition graph encoding music structures} \label{sec:DNA}

The problem we consider, i.e. how to restore a piece of music when an extended, 
connected piece has been lost or corrupted, often requires us to abandon the 
idea of exact recovery. In the case where only a short segment (up to about 
$50$ms) has been lost~\cite{adler2012audio}, or the signal can be described by a 
very simple structure~\cite{lagrange2005long}, it may be possible to infer the 
missing information from the regions directly adjacent to the distortion with 
sufficient quality. However, for complex music signals and corruptions of longer 
duration, such inference remains out of reach. Instead, we employ an analysis of 
the overarching medium- and large-scale structure of a music piece, determining
\emph{redundancies} in the signal to be exploited in the search for a replacement
for the distorted signal segment.

Conceptually, such analysis can be seen as a music segmentation
into chorus and verse, motifs and their variation, sections of 
equal or different meters, etc \cite{macpherson2008form}. The main difference to our approach is that, instead of 
working with high-level cognitive concepts such as meter and motifs, we instead consider
a basic time-frequency representation of the signal. In that representation, all the structures contained in a music recording are still preserved, although it is not always easily accessible to the human observer.

It is clear that repetition and less obvious redundancies do not occur to an equivalent
degree in every music piece. While they are an essential stylistic element to 
pop and rock music, certain movements, e.g. in contemporary music, attempt the
active avoidance of the familiar.
But even if a pattern is not repeated in the 
exactly same fashion, the conscious variation of previous structures, rhythmic, 
harmonic or otherwise, is an integral part of most music. Note that the grade of 
self-similarity inside a single recording may vary greatly. 

Going back to the original problem of music restoration, it seems natural to 
exploit this type of \emph{redundancy} in the musical piece to be restored. 
The temporal evolution of 
spectral content provides a surprisingly suitable first approximation of musical features.
Inspired by this observation, we construct an audio similarity graph. The 
vertices of the graph represent small parts of musical content, while the edges 
indicate the similarity between the segments in terms of local spectral 
content. The crucial step towards good performance is the enforcement of 
temporal coherence. This is achieved by selecting transitions that persist over
time, i.e. similarity is not instantaneous, but present for some period of time.
%
%

\section{Method} \label{sec:algo}

%
The ultimate goal of this contribution
is to provide a means for autonomous concealment of signal defects with a 
duration of a few hundred to several thousand milliseconds. 
\change{Here, we assume that the position of the defects is already known.}
\emph{The restoration 
should sound natural and respect the overall structure of the signal under 
scrutiny.} For short distortions, this implies, to some degree, the recovery
of the lost information in the defective region. For long gaps and dynamic signals,
we argue that it is of much greater importance than the transitions between the 
reliable signal segments and the proposed restoration sound natural. The further 
away from the transition points we are into the restored region, the less important
exact recovery becomes versus the restoration \emph{making sense} in the signal 
context. Therefore, we suggest an analysis of the signal structure with the proposed
similarity graph, to determine the most natural fit for the distorted region from
unaffected portions of the signal. The resulting method is an abstracted and 
autonomous version of manual restoration by searching the reliable signal for 
a replacement for the defective region. Since the proposed method forgoes the
synthesis of new audio content and provided that enough reliable signal content is available, the proposed method can handle signal defects of arbitrary length without affecting audio quality.

We obtain from a short-time Fourier transform~\cite{allen1977unified,Grochening:2001a,ga46} simple similarity features carrying important temporal and spectral information. On the basis of these features, a similarity graph is constructed, representing the temporal evolution and structure of 
the signal. If some signal segment is known to be defective, it is now sufficient to 
determine another segment of similar length, such that the beginning and end of the 
substitute resemble the signal before and after the defect. By placing the candidate 
segment at the previously corrupted position, the defect can be concealed.

\begin{figure}[htb!]
\begin{center}
\includegraphics[width=0.95\linewidth]{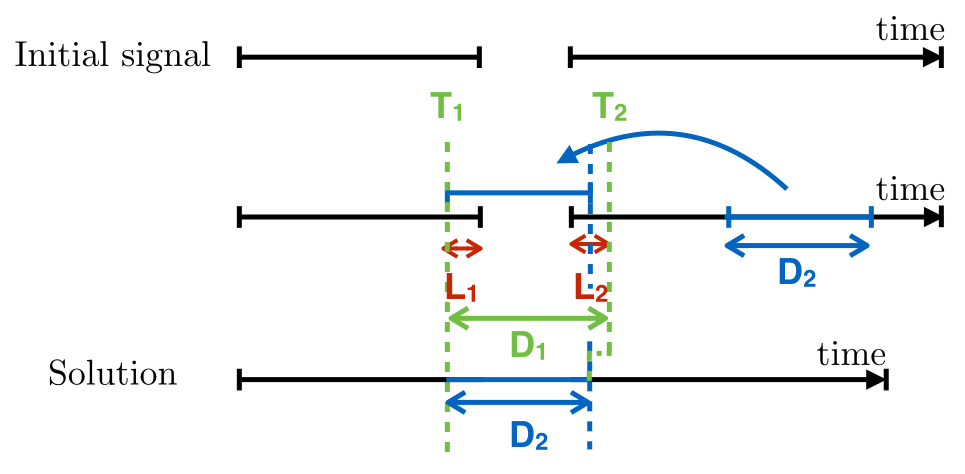}
\end{center}
\caption{\textbf{Illustration of the proposed inpainting method.} The determined candidate 
segment of duration $D_2$ is to be substituted for the gap. The optimal transition points
$T_1$ and $T_2$ are determined together with the candidate segment by jointly optimizing
(i) the similarity feature at $T_1$, $T_2$, (ii) the difference $|D_1-D_2|$ and the 
length of the necessary transition areas $L_1$ and $L_2$.}
\label{fig:main_idea}
\end{figure}

The proposed algorithm, illustrated in Figure~\ref{fig:main_idea}, searches for a replacement segment that optimally satisfies the three following criteria: 
\begin{enumerate}
\item The transitions $T_1$ and $T_2$ (light green dashed lines) resulting 
from the pasting operation should be perceptually transparent, i.e., the 
listener should not be able to notice the transition, even if the 
replacement segment does not correspond exactly to the missing data.
\item Some leeway is required for placing the transitions around the gap,
represented by $L_1$, $L_2$. However, the transition areas should not be 
unnecessarily long.
\item The length of the piece should remain approximately the same, i.e., 
the replacement duration $D_2$ should be close to the gap plus its surroundings, 
$D_1$. 
\end{enumerate} 
Some margin for compromise is, however, essential to the construction of a good 
solution. Since the question of how strictly the reliable content is to be 
preserved, i.e. how long $L_1$ and $L_2$ may be, is highly application-dependent, 
a parameter in the optimization scheme enables the tuning of this property.

In practice, at least for the inpainting problem, it is unnecessary to construct the full 
similarity graph. Consequently, we construct a sparsified graph which considers unique and
strong matches only. Weak matches are discarded. Only the strongest from a cluster of (temporally close) matches are considered. Finally, only edges connected to at least one node in the vicinity of the gap are relevant, since $L_1$ and $L_2$ are supposed to be small, see Figure \ref{fig:main_idea}. 

\subsection{Creation of the similarity graph}
The generation of the graph can be structured coarsely into $4$ distinct stages. In this section we disregard some technical details, instead concentrating on the general idea. The technical details of the individual steps of our method can be found in Sections~\ref{sec:TecDecGraph} and~\ref{sec:TecDecInp}. 

\begin{figure}[thb!]
\begin{center}
\includegraphics[width=0.95\linewidth]{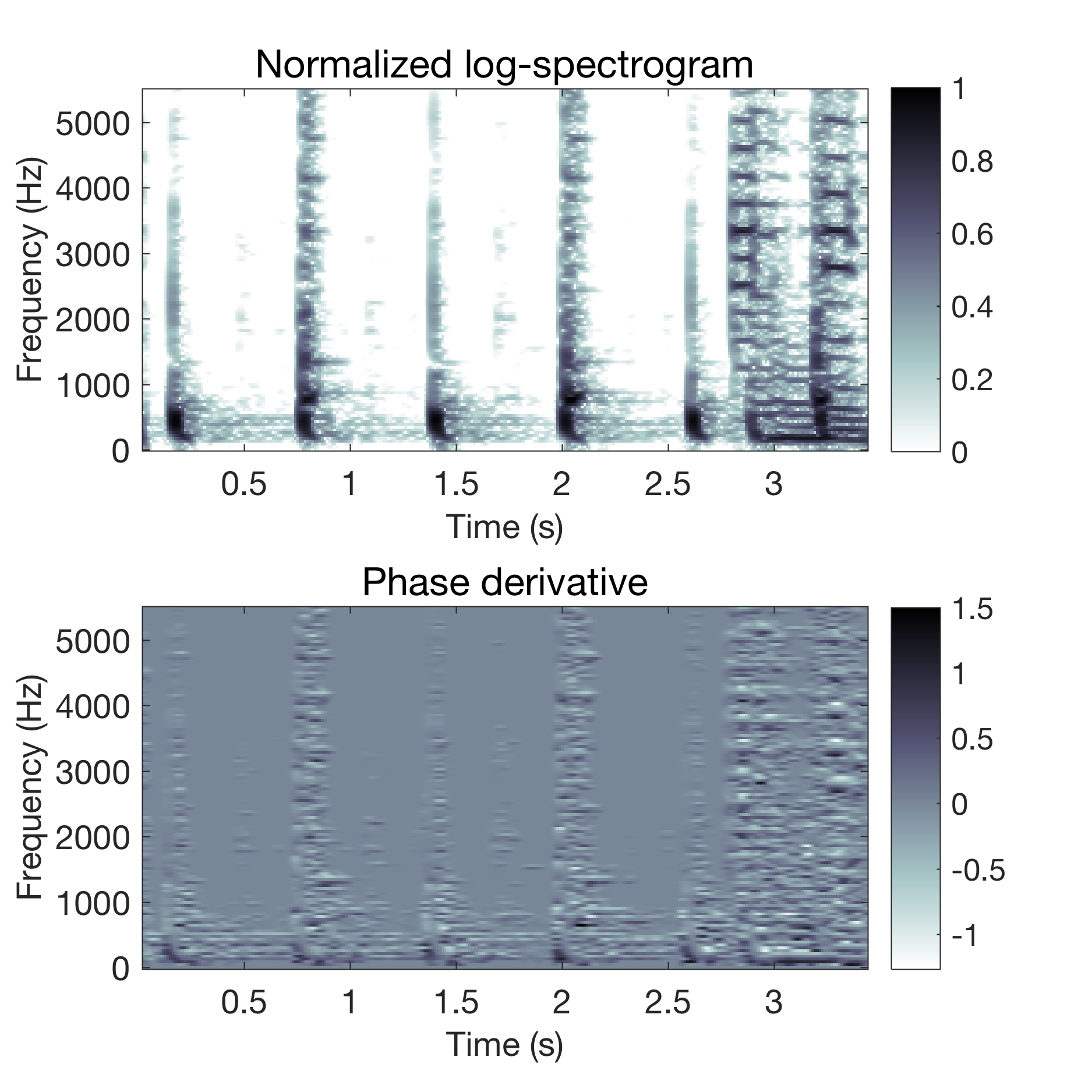}
\end{center}
\caption{Local audio features for an exemplary audio signal. The log-spectrogram $\tilde{F}_1$
(top) encodes the time-dependent intensity of frequency components. The smoothed partial phase derivative $\tilde{F}_2$ (bottom) has large values 
in the area of stable, harmonic components, independent of the component magnitude.\label{fig:features}}
\end{figure}

\paragraph{1. Compute basic similarity features} 
To determine temporal similarities in a signal, we have to settle on a \emph{feature}
that encodes the local signal behavior and a \emph{distance measure} that allows the 
comparison of feature vectors. For simplicity, and because the results were comparable
to more sophisticated features, we settle here on a weighted combination of two features
obtained directly from a short-time Fourier (STFT) analysis of the signal. Let $\bd C$ be
the matrix of short-time Fourier coefficients, with $\bd C_{m,n}$ denoting the coefficient obtained at 
the $n$-th time position in the $m$-th channel. $\bd C_{m,n}$, see Section \ref{sec:TecDecGraph}, can be decomposed uniquely into its magnitude $\bd M_{m,n}\geq 0$ and phase $\bd \Phi_{m,n}\in]-\pi,\pi]$ as 
\[
 \bd C_{m,n} = \bd M_{m,n}e^{i\bd \Phi_{m,n}}.
\]
Since the human auditory system perceives loudness approximately as a logarithmic function of sound pressure, the first part of our proposed feature is essentially a time slice of the dB-spectrogram, i.e. \[
 \tilde{F}_n^1 := [ 20\log_{10}(\bd M_{0,n}),\ldots,20\log_{10}(\bd M_{M-1,n})].
\]
\change{Note that direct spectrogram features have already proven to be useful in other applications, e.g. repetition-based source separation, see~\cite{rafii2013repeating}.

Additionally,} the time-direction partial derivative of the phase provides an estimate of the local instantaneous frequency~\cite{augfla95:reassign,Holighaus:2016:RSG:2910117.2910282}. \change{Let $\bd \Phi^{td}$ denote the $M\times N$-matrix containing the values of the time direction partial derivative of the phase, i.e. $\bd \Phi^{td}_{m,n} = \partial \bd \Phi_{m,\cdot}[n]$. The second part of our proposed feature is essentially
\[
  \tilde{F}_n^2 := \left[\bd \Phi^{td}_{0,n},\ldots,\bd \Phi^{td}_{M-1,n}\right],
\]
and $\tilde{F}_n = [\tilde{F}^1_n,\tilde{F}^2_n]^T$.} While $\tilde{F}^1$ puts a strong emphasis on signal components of high amplitude, $\tilde{F}^2$ attains large values for sinusoidal, or slowly frequency-varying, components independent of their magnitude, see also Figure \ref{fig:features}. This second part of the feature serves to emphasize low amplitude harmonic components, which may be highly important for perceived similarity. \change{The actual feature $\bd f_n$, defined in Section \ref{sec:TecDecGraph}, is conceptually equivalent to $\tilde{F}_n$, but implements some additional scaling.} Locality of the features is implied by obtaining the features from a STFT. The distance between two features at $l,k$ is simply the squared Euclidean distance of $\bd f_l$ and $\bd f_k$. 

\paragraph{2. Create a preliminary similarity graph}
The full (unprocessed) similarity graph determined from the given feature vectors would simply have all the time positions $n\in\{0,\ldots,N-1\}$ as vertices and edges connecting each vertex to every other vertex, with the associated weights derived from the distance between the associated features. 

\begin{figure}[thb!]
\begin{center}
\includegraphics[width=0.32\linewidth]{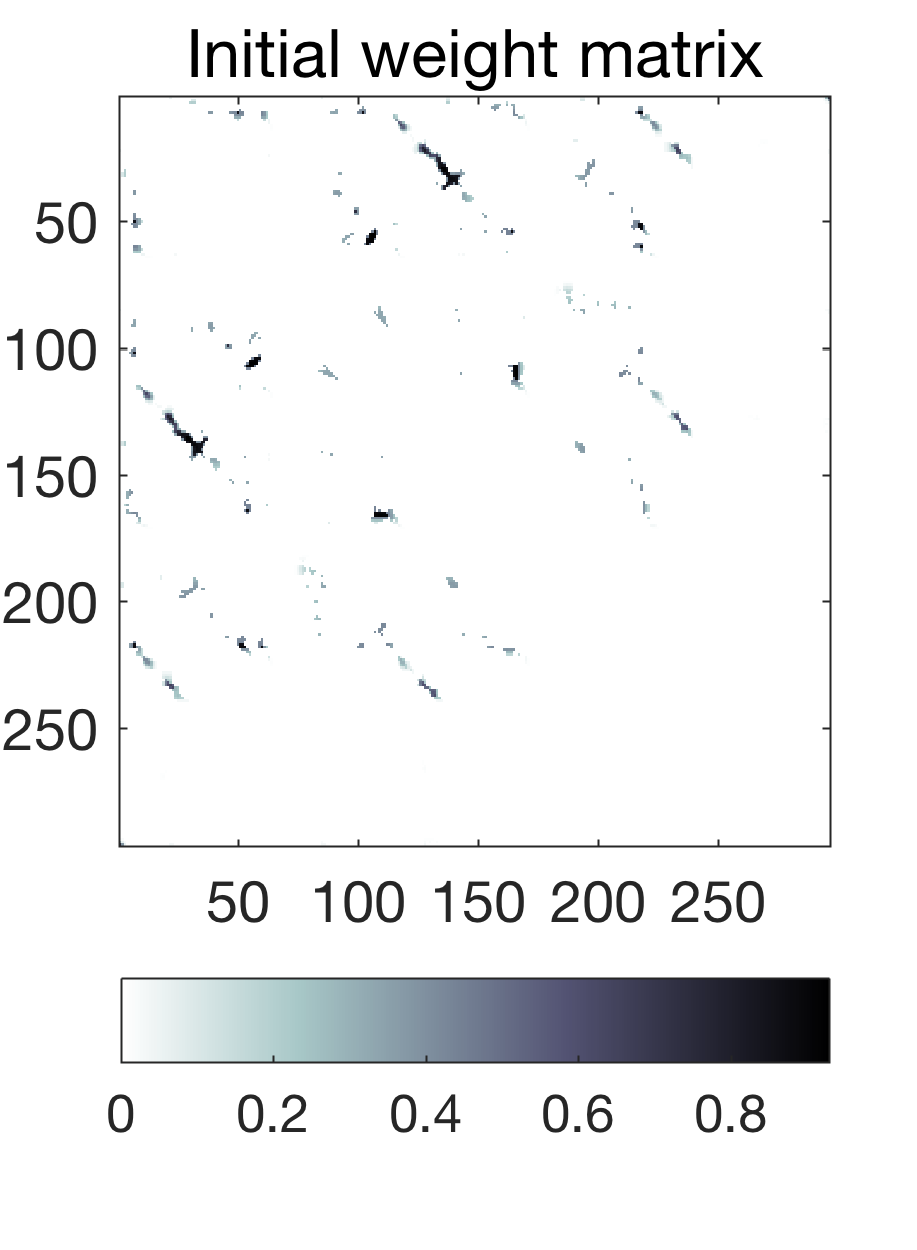}
\includegraphics[width=0.32\linewidth]{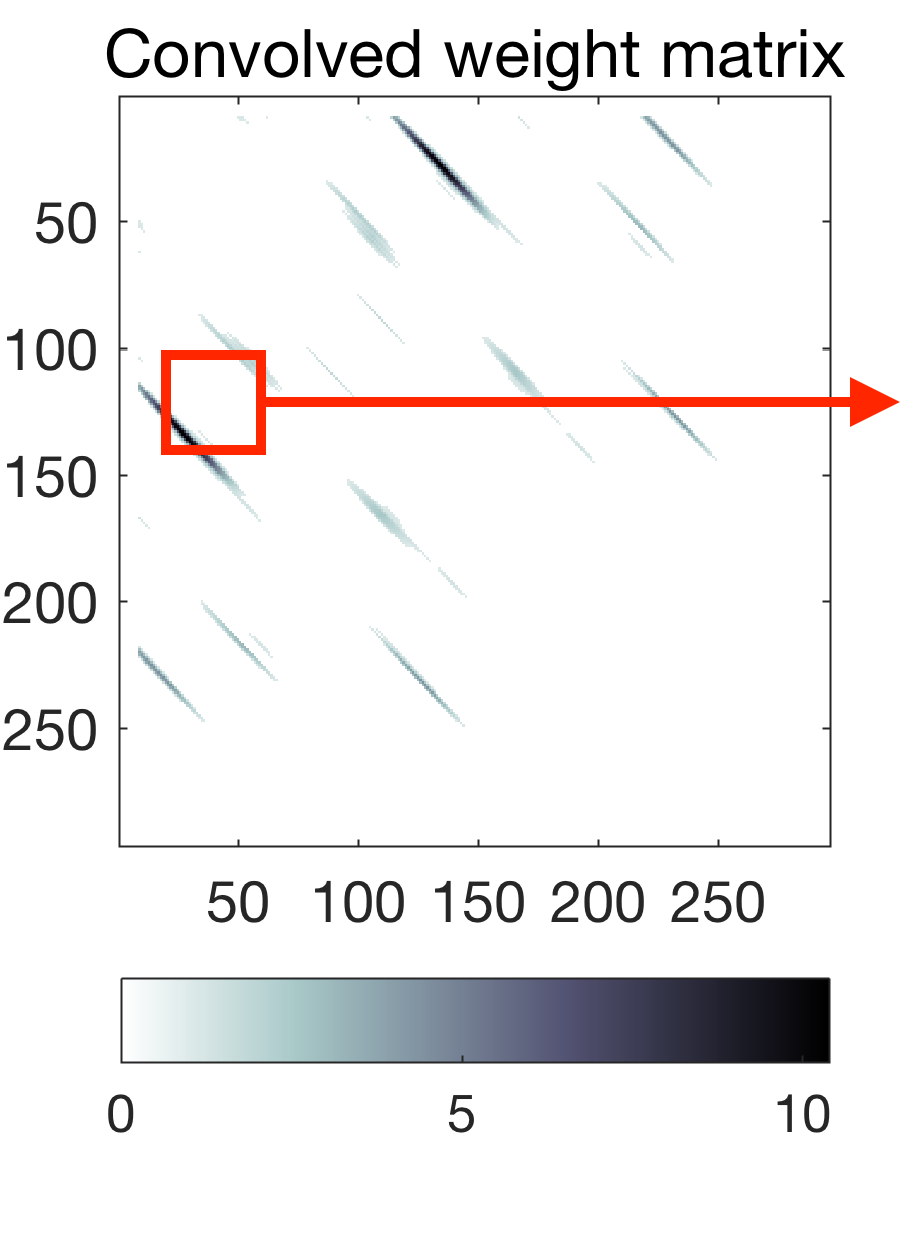}
\includegraphics[width=0.32\linewidth]{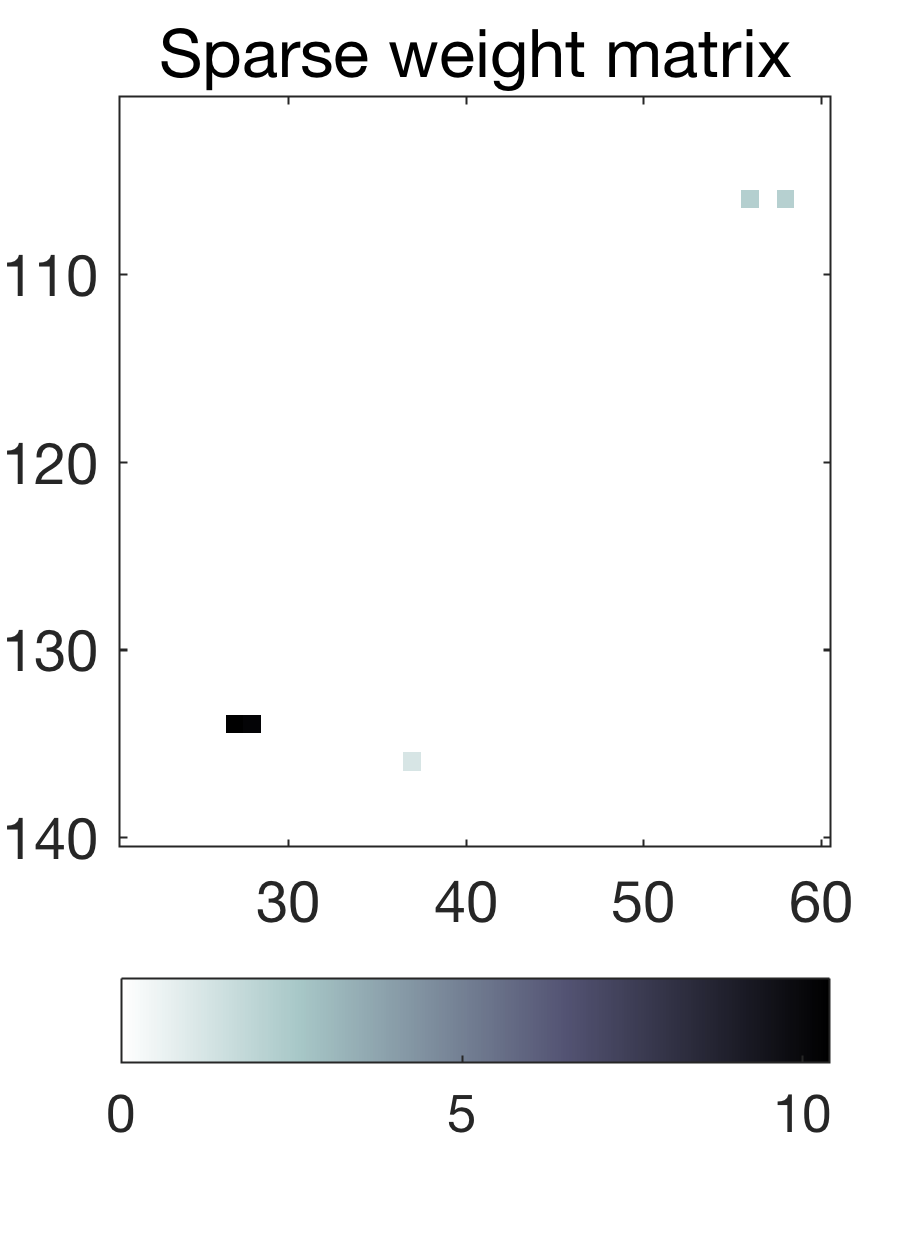}
\end{center}
\caption{Weight matrix based on feature vectors calculated for an exemplary 
audio signal \textit{without} a gap. Left panel: Preliminary weight matrix, 
$\bd{W_0}$, of the initial graph. Center panel: Convolved weight matrix, $\bd{W}$. 
Right panel: Excerpt of the weight matrix, $\bd{W_s}$, of the sparsified graph.}\label{fig:adjmatrices}
\end{figure}

The creation of such a graph is not only very expensive, but we are further only interested in a small number of strongest connections for every vertex. Therefore, we only determine the $K$ nearest neighbors, in terms of feature distance. Since this operation is expensive, we use the FLANN library (Fast Library for Approximate Nearest Neighbors)~\cite{muja2014scalable} to efficiently provide an approximate solution. For the $K$ determined neighbors, the edge weights are recorded in the adjacency matrix as \change{
\begin{equation}\label{eq:Wnull}
\bd{W_0}(l,k) = \begin{cases}
e^{-{\textstyle \frac{\|\bd f_l - \bd f_k\|_2^2}{\sigma}}} & \text{if } k \text{ is among the } K \text{ n.n.s of } l\\
0 & \text{otherwise,}\\
\end{cases}
\end{equation}}
for some $\sigma>0$, following a traditional graph construction scheme, see also Figure \ref{fig:adjmatrices} (left).

\paragraph{3. Enhance time-persistent similarities}
The individual features obtained from the STFT usually characterize signal's properties on a local time interval and do not capture the long-term signal's spectral characteristics. In order to capture longer temporal structures of a signal, we refine the graph by emphasizing its edges whenever a sequence of features at consecutive time positions is similar to another. In practice, 
this is achieved by convolving \change{the weight matrix $\bd{W_0}$} with a diagonal kernel $\bd D\in \RR^{L_K+1\times L_K+1}$, for some $L_K\in 2\NN$, with 
\[
 \bd D_{l,l} = 1-\frac{|L_K-2l|}{L_K} \text{ and } \bd D_{l,k} = 0, \text{ if } l\neq k.
\]
The resulting adjacency matrix is given as 
\begin{equation} \label{eq:weight_convolution}
\begin{split}
\bd W(l,k) & = (\bd{W_0}\ast \bd D)(l,k)\\ 
& = \sum_{l_0 =-L_K/2}^{L_K/2} \left(1-\left|\frac{2l_0}{L_K}\right|\right) \bd{W_0}(l+l_0,k+l_0),
\end{split}
\end{equation}
see also Figure \ref{fig:adjmatrices} (middle). Note that, in order to obtain an $N\times N$-matrix $\bd W$ and for the above equation to be valid $\bd W_0$ is implicitly extended to an $(N+L_K)\times(N+L_K)$-matrix with $L_K/2$ zeros on any side.

\begin{figure}[thb!]
\begin{center}
\includegraphics[width=0.475\linewidth]{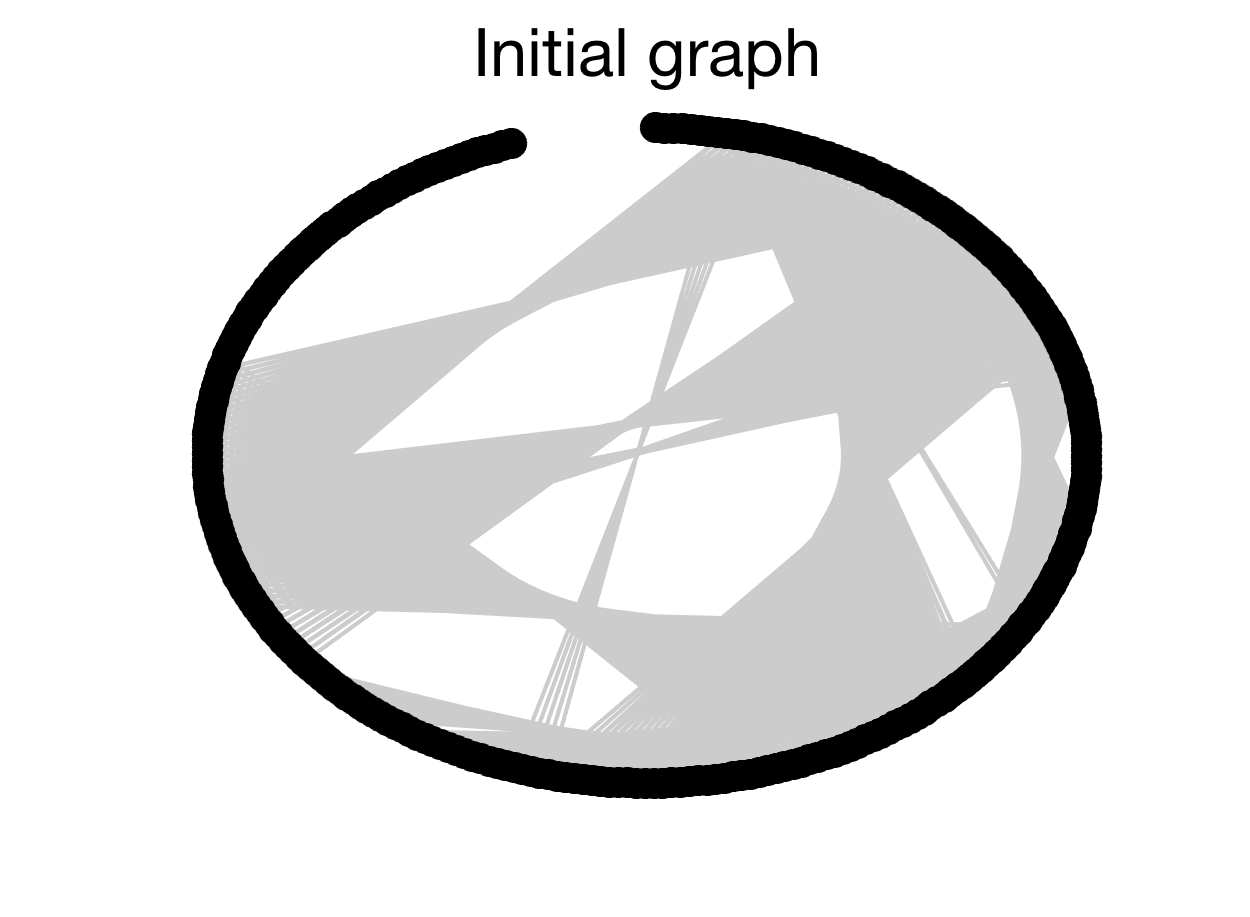}
\includegraphics[width=0.475\linewidth]{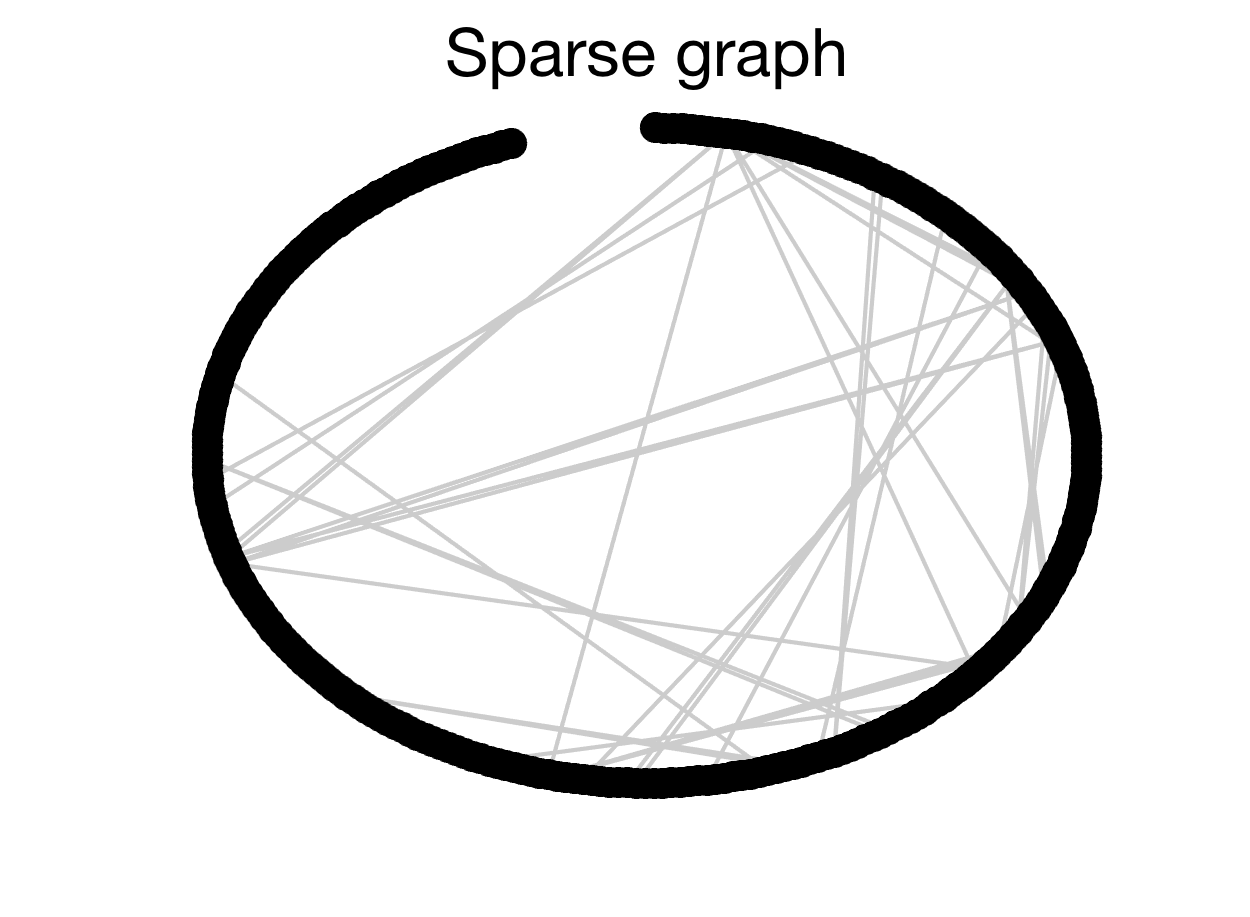}
\end{center}
\caption{Graphs based on feature vectors calculated for an exemplary audio 
signal \textit{without} a gap. \change{Left panel: Initial non-sparse graph, $G_0$, corresponding 
to the weight matrix $\bd{W_0}$, shown in Figure~\ref{fig:adjmatrices} (left). Right 
panel: Sparse graph, $G_s$ (only local maximum weights above the threshold 
considered), corresponding to the weight matrix $\bd{W_s}$, shown in 
Figure~\ref{fig:adjmatrices} (right).}}\label{fig:graphs}
\end{figure}

\paragraph{4. Delete insignificant similarities/Merge clustered similarities} 
After the convolution with the diagonal kernel, \change{the weight matrix $\bd{W}$} of our graph has been populated with a large number of nonzero entries, clustered around the entries of $\bd{W_0}$. The maxima of such clusters represent the strongest similarities between two regions of the signal. Moreover, only strong connections indicate significant similarities. Therefore, we delete all edges with weights below a certain threshold $t_w$ and select from every cluster of connections only the strongest, i.e the one with \emph{locally} the largest weight. \change{ This last step leads to the weight matrix $\bd{W_s}$ which is associated to the graph we use for our inpainting algorithm.} For an example of the final, sparsified adjacency matrix, see Figure~\ref{fig:adjmatrices} (right). Figure~\ref{fig:graphs} shows the difference between the original graph after Step 2 and part of the refined graph after Step 4. 

\subsection{Application: Audio inpainting and the reduced similarity graph} \label{sec:audio_inpainting_reduced}

The usage of the similarity graph for solving an inpainting problem is
rather straightforward. According to the paradigm described in Figure~\ref{fig:main_idea},
we want to find two edges \change{$(l_0,k_0)$ and $(l_1,k_1)$}, such that 
\begin{itemize}
 \item $l_0$ is close to the beginning of the distorted region and $k_1$ is close to its end,
 \item \change{$k_1-l_0$ is approximately equal to $l_1-k_0$} and
 \item \change{$\bd{W_s}(l_0,k_0)$ and $\bd{W_s}(l_1,k_1)$} are large. 
\end{itemize}
An appropriate choice of \change{$(l_0,k_0)$ and $(l_1,k_1)$} is determined by optimizing these $3$ criteria over all possible choices, for $l_0$ and $k_1$ in some limited range around the signal defect. The signal segment corresponding to the local features \change{$k_0,\ldots,l_1$} is then substituted for the original signal in the range corresponding to $l_0,\ldots,k_1$. 

\begin{figure}[thb!]
\begin{center}
\includegraphics[width=0.45\textwidth]{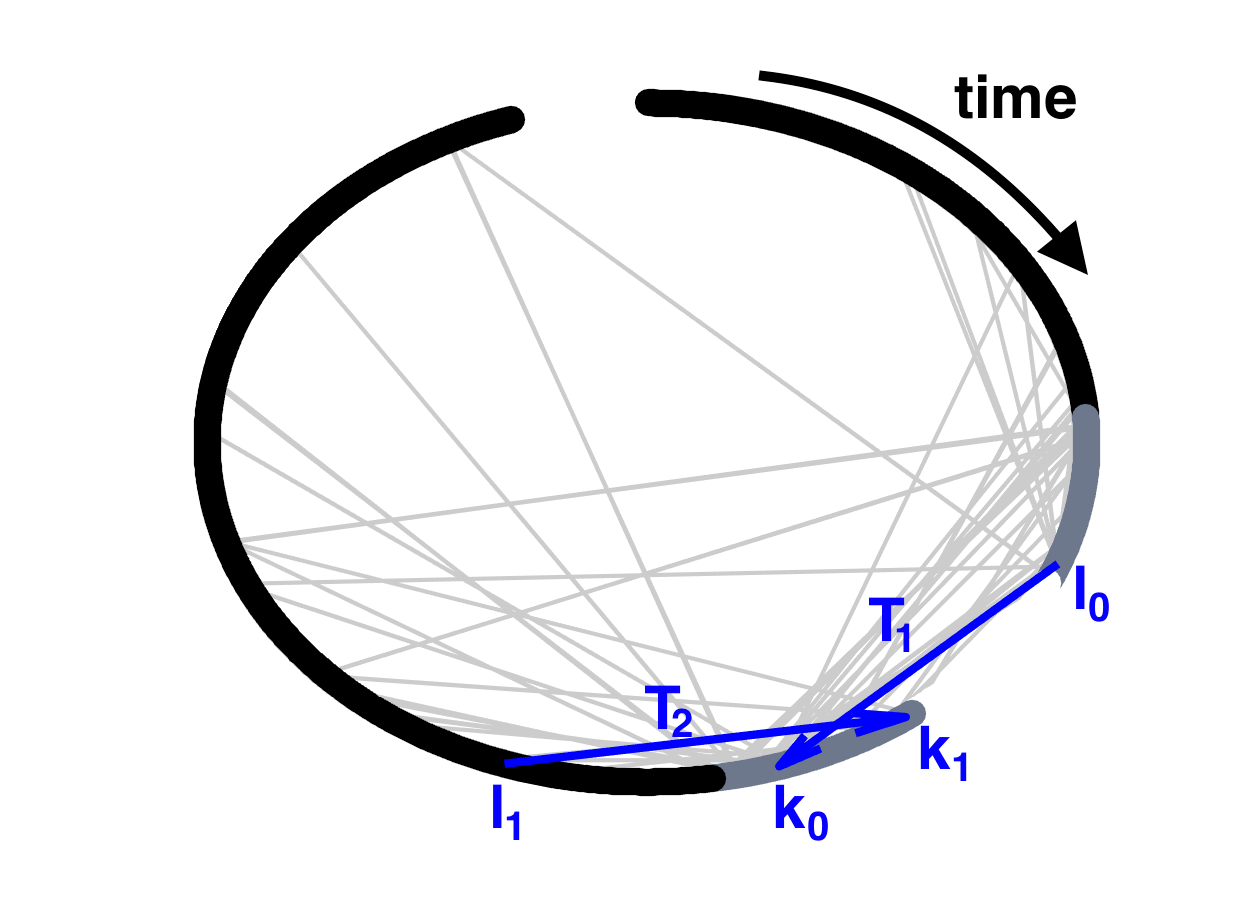}
\end{center}
\caption{Final reduced graph based on exemplary audio features calculated from an audio 
signal \textit{with a gap}. \change{The 
regions considered for the transitions are in gray with the gap in between them in white. All available transitions for the reconstruction are in light gray with the optimally selected $T_1$ and $T_2$ in blue. The nodes indexes $l_0,l_1$ and $k_0,k_1$ correspond the beginnings and the ends of the transitions $T_1$ and $T_2$} }\label{fig:subgraph}
\end{figure}

For the purpose of inpainting, we are only interested in edges that connect to at least one vertex either
shortly before, or shortly after, the signal defect. Hence, only a small horizontal (or vertical) slice of the sparse matrix $\bd{W_s}$ has to be computed, greatly reducing the complexity of the graph creation. \change{Figure~\ref{fig:subgraph} shows an example of such a reduced graph (not to be confused with the sparse graph) and the determined transitions \change{$T_1$ indexed by $(l_0,k_0)$ and $T_2$ indexed by $(l_1,k_1)$} for an exemplary signal and defect. In practice we use the reduced for graph all experiment of this paper.}

\section{The similarity graph in detail}\label{sec:TecDecGraph}

\subsection{Local audio features} \label{sec:audio_features}
\change{Building a similarity graph for full music pieces from STFT features is in practice challenging simply due 
to the number and size of the obtained features. }To be efficient, the number of features has to remain small in contrast to the complexity of audio signals. Our solution leverages two techniques to obtain a good trade-off: 1) an adequate sub-sampling, and 2) a tight low-redundancy STFT.

While audio signals are often sampled at a very high rate, to compute 
reliable audio features, a much lower rate is usually sufficient. We choose 
a maximum sampling rate of $\xi_{\text{max}}$~Hz (default $12$~kHz, see Table~\ref{tab:parameters}
for all default parameters). If a given signal $s\in\RR^L$ is sampled at a higher rate $\xi_s$~Hz, $s$ is decimated with a decimation factor $d=\lceil \xi_s/\xi_{\text{max}} \rceil$, after the application of 
an anti-aliasing filter. We denote the decimated signal by $s_d$.

The short-time Fourier transform (STFT) of $s_d$ with respect to a (real-valued) window function $g$,
hop size $a\in\NN$ and $M$ channels is 
defined as
\[
 \bd C_{m,n}:= \sum_{l=0}^{L-1} s_d[l]g[l-na]e^{-2\pi iml/M}, 
\]
for $n\in\{0,\ldots,L/a-1\}$ and $m\in\{0,\ldots,M-1\}$. Recall the decomposition of $\bd C_{m,n}$ into magnitude and phase: $\bd C_{m,n} = \bd M_{m,n}e^{i\bd \Phi_{m,n}}$, $\bd M_{m,n}\geq 0$, $\Phi_{m,n}\in]-\pi,\pi]$. By default, we choose $g$ to equal a $1024$-point Itersine window~\cite{wesfreid1993adapted}, $a=128$ and $M=1024$. This particular construction leads to an $8$ redundant tight frame, hence preserving equally each signal frequency component. 

The $2$ separate parts $\bd F^1_n$ and $\bd F^2_n$ of the feature vector $\bd f_n$ are obtained as follows.

\paragraph{dB-Spectrogram} Let $\bd S^{dB}_{m,n} := 20\log10(\bd M_{m,n})$, $n\in\{0,\ldots,L/a-1\}$ and $m\in\{0,\ldots,M-1\}$. For more convenient handling, $\bd S^{dB}$ is limited to a fixed range and peak-normalized, resulting in 
\[
  \bd F^{1}_{m,n} = t_s^{-1}\left(\bd S^{dB}_{m,n}-\max_{k,l}(\bd S^{dB}_{k,l})+t_s\right)_{+},
\]
where $(x)_+ = x$, if $x>0$, and $0$ otherwise. By default, $t_s=50$~dB. Figure~\ref{fig:features} (top) shows $\bd F^{1}$ for an exemplary audio signal.

\paragraph{Relative instantaneous frequency} In~\cite{augfla95:reassign},
the authors show that an instantaneous frequency estimate can be associated to 
$C_{m,n}$ by 
\begin{equation}\label{eq:instFreq}
 \tilde{\xi}_{m,n} := \frac{\xi_sm}{M} -\mathbf{Im}(\bd C^{td}_{m,n}/\bd C_{m,n}),
\end{equation}
where $\bd C^{td}_{m,n}:= \sum_{l=0}^{L-1} s_d[l]g'[l-na]e^{-2\pi iml/M}$ and $g'$ is a discrete derivative of $g$.
The second term in the equation above is in fact an equivalent expression for the 
partial derivative $\bd \Phi^{td}_{m,n}$ of $\bd \Phi$, with respect to $n$. $\tilde{\xi}_{m,n}$ might fluctuate quickly and its range depends on $m$. Both these properties are undesired for our purpose. 
Therefore, we consider only its relative part, i.e. the second term in Eq. \ref{eq:instFreq}, and
perform a channel-wise smoothing of each $\tilde{\xi}_{m,\cdot}$, $m\in 0,\ldots,M-1$, by convolution with a localized kernel $v_{\text{ker}}$ (default: $8$-point Hann window). 
Additionally, the expression for $\tilde{\xi}_{m,n}$ is unstable in regions of small magnitude $\bd M_{m,n}$ \cite{xxlbayjailsoend11}. With $t_p = \max_{m,n} |\mathbf{Im}(\bd C^{td}_{m,n}/\bd C_{m,n})|$, we define 
\[
  \bd F^{2}_{m,n} = \begin{cases} 
                      -t_p^{-1}\left(\mathbf{Im}(\bd C^{td}_{m,\cdot}/\bd C_{m,\cdot})\ast v_{\text{ker}}\right)[n] & \text{ if } \bd F^{1}_{m,n}>0,\\
                      0 & \text{ else.}
                    \end{cases}
\]

The combined feature vector is obtained as 
\begin{eqnarray*}
 &\bd f_{n} = ( \bd F^{1}_{1,n},\ldots,\bd F^{1}_{M-1,n},\lambda\bd 
F^{2}_{1,n},\ldots,\lambda\bd F^{2}_{M-1,n})^T,
\end{eqnarray*}
for $n=0,\ldots,L/a-1$. \change{We choose a default value of $\lambda = 3/2$, since this choice resulted in similar importance placed on both sub-features.}

\subsection{Creation of the similarity graph} \label{sec:similarity_graph}
When it comes to the graph creation, we desire an automatic parameter selection adapting to the audio features.
For the creation of the initial graph, we only need to determine the value of $\sigma$ in 
the expression \eqref{eq:Wnull} for the preliminary weight matrix. Denoting as $K_n$ the 
set of $K$ approximate nearest neighbors of the vertex $n$, our solution is to set 
$\sigma$ to the average squared nearest neighbor distance
\[
 \sigma = \frac{1}{NK}\sum_{n=0}^{N-1}\sum_{l\in K_n} \|\bd f_n - \bd f_l\|_2^2.
\]
Thus $\bd W_0(l,k)\approx 1$ if $\bd f_l$ and $\bd f_k$ are close, and decreasing towards $0$, the more
$\bd f_l$ and $\bd f_k$ differ. Our experiments showed that $K$ of $40$ is a good default value, which should be increased if the music is expected to be very redundant.

\begin{figure}[htb!]
\begin{center}
\includegraphics[width=0.475\linewidth]{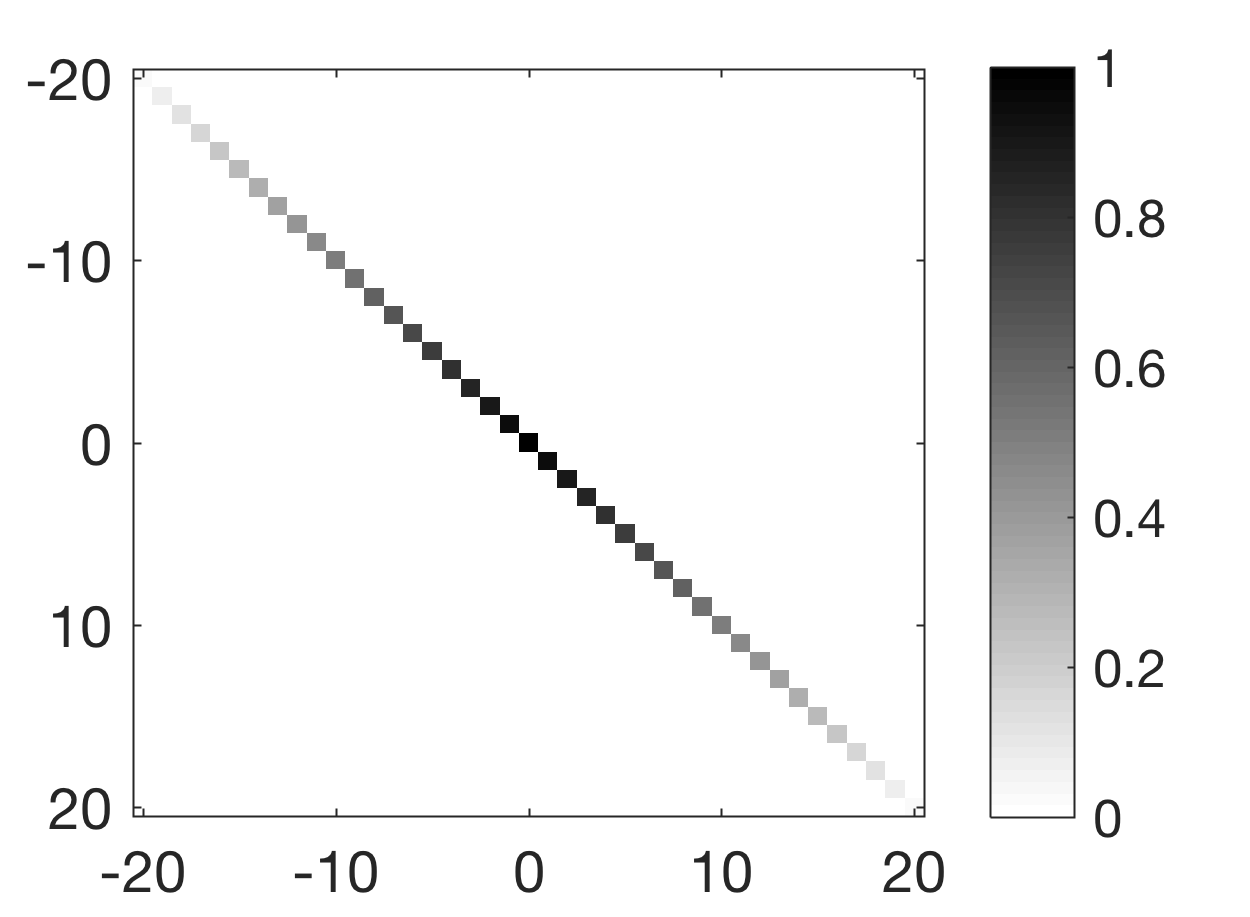} 
\end{center}
\caption{Convolution kernel used to enhance the diagonal shape of the weight 
matrix. \change{Here $L_k=40$.} \label{fig:kernel}}
\end{figure}

To obtain $\bd{ W}$ from $\bd{W_0}$ in~\eqref{eq:weight_convolution}, the length of the convolution kernel must be fixed. After 
the convolution, the edges in the graph describe the similarity of 
 signal segments
of $\frac{aL_K}{\xi_s}\lceil \xi_s/\xi_{\text{max}}\rceil$~seconds duration. The choice of 
$L_K$ determines the importance of \emph{long duration similarities} over such with short duration.
We used \change{$L_K = 40$} as a default value in order to consider roughly half-second segments for signals
sampled at $44.1$~kHz, see Figure \ref{fig:kernel}.

To transition from $\bd W$ to $\bd W_s$, we first perform a thresholding
by $t_w$. In $\bd{W_0}$, each entry can be 1 at maximum, see \eqref{eq:Wnull}. In $\bd{ W}$, solitary entries will be smaller than 1 and entries surrounded by other high-valued entries will be larger than 1. In order to suppress solitary entries, we used $t_w = 2$ as a default value. 
The final step consisting of selecting the local maxima by choosing points that are equal to or larger than the four direct neighbors.
\change{In detail $\bd{W_s}$ is defined as
\begin{equation*}
\bd{W_s} (l,k)= \begin{cases} 
\bd{W}(l,k) &  \text{if } \bd{W} (l,k) \geq \\
& \indent \max\{ t_w, \bd{W} (l\pm 1,k\pm 1) \}\\
0  & \text{otherwise.}
\end{cases}
\end{equation*}
The notation $\pm$ should be interpreted as the collection of all possible choices, i.e. the maximum over all direct neighbors in $\bd{W}$
}

When applying the calculation of the transition graph to a signal where the 
distorted area is known, the computational cost can be further reduced. 
In particular, only a partial 
transition graph needs to be computed because we are only interested in outgoing 
connections within the short region before, and incoming connections 
within a short region immediately after the distortion. Conceptually, we consider only 
small $L_1$ and $L_2$, cp. Figure~\ref{fig:main_idea} and Section \ref{sec:transitions}. \change{Therefore, the $K$ 
nearest neighbors search and all the following operations, is not performed on all nodes, but only for a small 
subset of features in the direct vicinity of the signal defect. This allows us to not only greatly reduce the computation cost, but also reduce the size of the optimization problem described in Section~\ref{sec:transitions}. An example of such resulting graph is given in Figure~\ref{fig:subgraph}.}

\section{The inpainting step in detail}\label{sec:TecDecInp}

\subsection{Selection of optimal transitions} \label{sec:transitions}

To select the optimal transition, we need to transform the three conditions of Section~\ref{sec:audio_inpainting_reduced} into a mathematical objective function.
Let $d_s,d_e$ denote the 
index of the nodes corresponding to the start and end of the distorted region. \change{In the notation 
of the previous section, only edge $(l_0,k_0),(l_1,k_1)$ with $l_0\in L_1:=\{[d_s-\epsilon_1,d_s]$
and $k_1\in L_2:=[d_e,d_e+\epsilon_2]\}$ are considered acceptable. In our experiments, we observed that setting $\epsilon_1=\epsilon_2$ to a length corresponding to approximately 5 seconds yielded good results. The region considered for possible transition can be seen as the red interval in Fig \ref{fig:subgraph}.}

Among all acceptable edges, we search for the solution that 
minimizes the \emph{objective function}
\change{\begin{equation}
\begin{split}
f\left((l_0,k_0),(l_1,k_1)\right) & = \left|(k_1-l_0)-(l_1-k_0)\right| \\
 & + \gamma_2 \left((d_s-l_0) + (k_1-d_e) \right) \\
 & + \gamma_3 \left( \frac{1}{\bd{W_s}(l_0,k_0)} + \frac{1}{\bd{W_s}(l_1,k_1)}\right).
 \end{split}
\end{equation}}
Compare the definition of $f$ with Figure \ref{fig:main_idea} to see that:
The first term controls the difference $D_2-D_1$, the second term the distances
$L_1,L_2$ from the defect and the third term controls the quality of the transitions.
By tuning $\gamma_2$ and $\gamma_3$, we can vary the importance of the individual terms.
In our experiments, $\gamma_2 = 1$ and $\gamma_3 = 100$ have provided good results.

Since the number of acceptable transitions is small, the computational benefit from 
using a sophisticated optimization algorithm is negligible. Hence, we 
solve the optimization problem by simply computing exhaustively the values of the 
objective function for each set \change{$(l_0,k_0),(l_1,k_1)$ with $l_0\in L_1$
and $k_1\in L_2$.}

\subsection{Signal reconstruction} \label{sec:reconstruction}
When two audio signals are concatenated naively, discontinuities and phase jumps might 
result in clicking artifacts. To reduce these effects, a smoothed
transition is clearly preferred. We propose the following: 
Since the features are obtained from a STFT with time step $a$, with respect 
to a possibly decimated signal, the \emph{time resolution} of the similarity 
graph analysis equals \change{$\tilde{a}:= a\lceil \xi_s/\xi_{\text{max}}\rceil$} samples.
In other words, the preliminary solution obtained in the previous step suggests
the insertion of the signal samples \change{$s[\tilde{a}k_0,\ldots,\tilde{a}l_1-1]$} in 
place of \change{$s[\tilde{a}l_0,\ldots,\tilde{a}k_1-1]$}. 
To further improve the transition, we \change{allow to adjust the transition positions 
$\tilde{a}l_0$ and $\tilde{a}k_1$ by up to half the similarity graph's time resolution, i.e. $\tilde{a}/2$ 
samples. The optimal adjustment is determined by maximizing a correlation, as 
proposed in \cite{bahat2015self} and described below.} Denote by $L_w$ the length of the 
analysis window
$g$ and \change{$\tilde{L}_w:= \lceil L_w/2\rceil \lceil \xi_s/\xi_{\text{max}}\rceil$}. 
The final transitions are given 
by \change{$(\tilde{l}_0,k_0),(\tilde{a}l_1,\tilde{k}_1)$},
where \change{
\[
\begin{split}
  \tilde{l}_0 & = \argmax_{l\in{[}\tilde{a}l_0-\tilde{a}/2,\tilde{a}l_0+\tilde{a}/2[} \langle s_{l},s[\tilde{a}k_0-\tilde{L}_w,\ldots,\tilde{a}k_0+\tilde{L}_w-1]\rangle,\\
  \tilde{k}_0 & = \argmax_{l\in{[}\tilde{a}k_1-\tilde{a}/2,\tilde{a}k_1+\tilde{a}/2[} \langle s_{l},s[\tilde{a}l_1-\tilde{L}_w,\ldots,\tilde{a}l_1+\tilde{L}_w-1]\rangle.
\end{split}
\]
Here, $s_l\in\RR^{2\tilde{L}_w}$ is the vector 
\[
 s_{l}[j] = \begin{cases} 
                0 & \text{ if } l-\tilde{L}_w+j \in[d_s,d_e],\\
                s[l-\tilde{L}_w+j] & \text{ otherwise.}
              \end{cases}
\]}
The obtained indices $\tilde{l}_0,\tilde{k}_1$ maximize the correlations between the original signal and the inpainting candidate.

\change{In order to obtain smooth transitions in the restored signal, we perform a time-frequency domain 
cross-fading. Conceptually, this requires us to consider $3$ different arrays of short-time Fourier coefficients
with time step offsets $\tilde{l_0}-l_0\tilde{a}$, $0$ and $\tilde{k_1}-k_1\tilde{a}$, respectively:
\[
\begin{split}
 \bd C^{(1)}_{m,n} & = V_{\tilde{g}} 
\tilde{s}[(n-l_0)\tilde{a}+\tilde{l}_0,mL/M], \\ 
 \bd C^{(2)}_{m,n} & = V_{\tilde{g}} 
\tilde{s}[n\tilde{a},m\tilde{L}/\tilde{M}], \\ 
 \bd C^{(3)}_{m,n} & = V_{\tilde{g}} 
\tilde{s}[(n-k_1)\tilde{a}+\tilde{k}_1,mL/M].
\end{split}
\]
This ensures that in $C^{(1)}$, the $l_0$-th time frame is centered at the signal position $\tilde{l_0}$ and
in $C^{(3)}$, the $k_1$-th time frame is centered at position $\tilde{k_1}$. The analysis window
$\tilde{g}$ is chosen, such that on the undecimated signal $s$, it mimics $g$ acting on $s_d$. Hence, its length 
and the number of channels $M$ are set to $2\tilde{L}_w$. Thus, by default, we choose $\tilde{g}$ to also be of Itersine shape.}

The restored signal can now be obtained by applying the inverse STFT 
to the combined matrix,
\[
\begin{split}
 \bd C^{\text{rec}} & = \Big(\bd C^{(1)}_{\cdot,1},\ldots,\bd C^{(1)}_{l_0-1}, 
\bd C^{(2)}_{\cdot,l_1},\ldots,\bd C^{(2)}_{\cdot,k_0-1}, \\
 & \hspace{90pt} \bd C^{(3)}_{\cdot,k_1},\ldots,\bd C^{(3)}_{\cdot,L/\tilde{a}-1} 
\Big).
\end{split}
\]			
In practice, complexity is further reduced without altering the result, by computing $\bd C^{(j)}$, $j=1,2,3$, only for the time-positions relevant to the cross-fading, thus obtaining two small submatrices of $\bd C^{\text{rec}}$. \change{Note that any coefficient vector $\bd C^{(j)}_{\cdot,n}$, $j=1,2,3$, $n=0,\ldots,N-1$, only affects the reconstruction on an interval equal to the window length $2\tilde{L}_w$. Hence, both transitions have a duration of $2\tilde{L}_w$ and the first can be recovered from 
\[
 \Big(\bd C^{(1)}_{\cdot,l_0-r},\ldots,\bd C^{(1)}_{l_0-1}, 
\bd C^{(2)}_{\cdot,l_1},\ldots,\bd C^{(2)}_{\cdot,l_1+r-1}\Big), 
\]
and similarly for the second transition. Here, $r:= 2\lceil \tilde{L}_w/\tilde{a}\rceil$ is a generous estimate of the ratio between the window length and the hop size.} The inverse STFT is then applied to these submatrices and the cross-fade regions, which are obtained as the central part of those inverse STFTs, are placed at the desired position in the signal. All other operations are performed in the time domain. To ensure equivalence with a complete STFT computation, the segments have to start/end $M$ samples before/after the cross-fading.  

\section{Numerical evaluations}\label{sec:experiments} \label{sec:evaluation}

In this section we provide a numerical evaluation of the proposed algorithm. 

First, we 
verify the algorithm in a setting where the gap content is provided with the 
remaining signal. A correct implementation should be able to perfectly replace 
the gap by exactly the lost content. Second, we investigate algorithm's computational 
performance in terms of average runtime. 


For the evaluations, the algorithm was implemented in MATLAB. The implementation 
is based on LTFAT \cite{ltfatnote030} for feature extraction, and on the GSPBox 
\cite{perraudin2014gspbox} for graph creation. For non-commercial use, the 
algorithm is available 
online\footnote{\change{\url{https://epfl-lts2.github.io/rrp-html/audio_inpainting/}}}, alongside a 
browser-based demonstration 
\footnote{\change{\url{https://lts2.epfl.ch/web-audio-inpainting/}}}. Table 
\ref{tab:parameters} provides a summary of the algorithm parameters used for the 
evaluations.

\subsection{Verification}\label{ssec:experimentSanity} 
Here, we address the question whether the algorithm perfectly recovers the gap 
when an exact copy of the missing segment is present within the reliable signal. 
For this purpose, we used a set of $16$ uncorrupted audio signals with various 
content and at the sampling rate of $44100$ Hz. First, redundant signals were 
created by repeating the signal, i.e. placing a copy of the signal at its end. 
Then, each redundant signal was corrupted by creating a gap of $2$ seconds. For 
each signal, the experiment was repeated five times with randomly chosen position of 
a gap, yielding $80$ corrupted signals. Then the algorithm was applied on each 
of the corrupted signal. \change{In all reconstructions, the $\ell^2$-norm difference 
between the original and reconstructed signals was in the range of numerical 
precision, implying that each corrupted signal was perfectly restored.}
Hence, we consider the implementation of the presented algorithm as verified. 

\subsection{Computational complexity}\label{ssec:experimentComp}
The algorithm can be separated into different steps that all have different computational requirements.
Here, we investigated the individual costs of each step and their relative importance in the overall performance of the algorithm.
The evaluation was performed on a \change{modern notebook ($2.5$ GHz Intel i7, 2 cores, $16$ GB RAM)} for the same set of corrupted signals \change{as in 
Sec.~\ref{ssec:experimentMain}}. Table \ref{tab:timing} shows mean and standard 
deviation of the computation time per minute of audio signal. On average, each 
minute of audio signal required $2.47$-s computation time for the reconstruction.

The feature computation, graph creation and the selection of the optimal 
transition\change{, performed on the reduced sparse graph (see Figure~\ref{fig:subgraph}), }
scale linearly with the length of the provided reliable data, in 
terms of both storage and time complexity. As a result, our result consists of the timing per minute of analyzed music.
In all our experiments, the reliable 
data was given by a full song, without the corrupted segment. 
\change{Note that linear complexity can only be achieved by considering the
reduced graph. For the full sparse graph $\bd W_0$, complexity of the graph creation
is $\mathcal O(N \log N)$ and the transition selection would even scale roughly quadratically, i.e. 
be $\mathcal O(N^2)$. Even if the selection is restricted to the range considered in the 
reduced graph, linear complexity would be out of reach. Therefore, the computation
time per minute is not a reliable indicator anymore. We just remark that on the dataset used,
the graph construction was on average $~8$ times slower, while the average duration for 
transition selection increased by a factor of $~40$, when performed on the full sparse graph.
Although we did not systematically evaluate memory usage of the method, it should be noted that 
restricting to the reduced graph is considerably more efficient in that regard, as well.}

If multiple 
corruptions are to be removed using the same set of reliable data, the algorithm 
benefits from the fact that features only need to be computed once. Since the 
feature computation is the bottleneck of the method (this can be seen in Table 
\ref{tab:timing}), this may lead to significant boosts of computational performance in the case 
of multiple gaps.

\begin{table}[ht!]
\begin{center}
\change{\begin{tabular}{|l|c|c|}
\hline 
{\bf Processing step} & {\bf Reduced graph} (Mean)  & {\bf Reduced graph} (STD)\\
\hline
\hline
Feature extraction & $1.84$ & $0.18$\\
\hline
Graph construction & $0.50$ & $0.06$\\
\hline
Transition selection & $0.02$ & $0.004$\\
\hline
Signal reconstruction & $0.04$ & $0.007$\\
\hline
\hline
{\bf Total} & $2.47$ & $0.18$ \\
\hline
\end{tabular}}
\end{center}
\caption{\label{tab:timing}\change{Average execution time of the proposed methods per minute of provided audio (Based on a database of $16$ songs) for the reduced graphs, see Fig.~\ref{fig:subgraph}.
}
}
\end{table}
\section{Perceptual evaluation}\label{ssec:experimentMain}

In order to estimate the potential of the proposed algorithm for music, we conducted a psychoacoustic test, in which we evaluated the impact of the artifacts occurring from inpainting various songs from a music database. In particular, we were interested in addressing the following questions:
\begin{enumerate}
	\item How often are subjects able to detect an alteration (detectability)? The answer gives us access to how often our algorithm is able to fool the listener.
	\item How precisely can subjects pinpoint the alteration? The answer gives us an indication of the inpainting quality and of the confidence of the test subject.
	\item How disturbing are the detected artifacts (severity)? The answer provides some good {insights into} the reconstruction quality even when the listener is not fooled.
	\item Is the familiarity of the song correlated with the detectability or the severity? The answer gives some {intuition about} the quality of the reconstruction and ensures that we are not only fooling the non-familiar test subjects.
\end{enumerate}

In order to ensure that our experiment provides meaningful results truly describing the potential of the proposed algorithm, our subjects were familiar with the tested music genres and we have collected ratings for familiarity and liking the songs.

\subsection{Testing methodology}
\paragraph{Material}
The sound material consisted of songs from the following genres: \change{pop, rock, 
jazz, classical. These genres were selected to cover the most common listening habits and, with respect to music structure also include many other, similar genres like blues, country, folk, oldies, hip-hop, etc}. Six songs per genre were selected from hundreds of songs with the 
aim to well-represent the genre. 

\paragraph{Subjects}
In order to test subjects familiar with our material, in a self-assessment questionnaire, \change{a candidate had to provide the average weekly listening duration (in hours) to the genres pop, rock, jazz, classical, and others}. For the evaluation, only candidates listening at least 4 hours per week to music from \change{all four main genres in total} were considered. In total, 15 subjects were selected for the test. They were paid on an hourly basis. 

\begin{figure}[htb!]
	\begin{center}
		\includegraphics[width=1.0\linewidth]{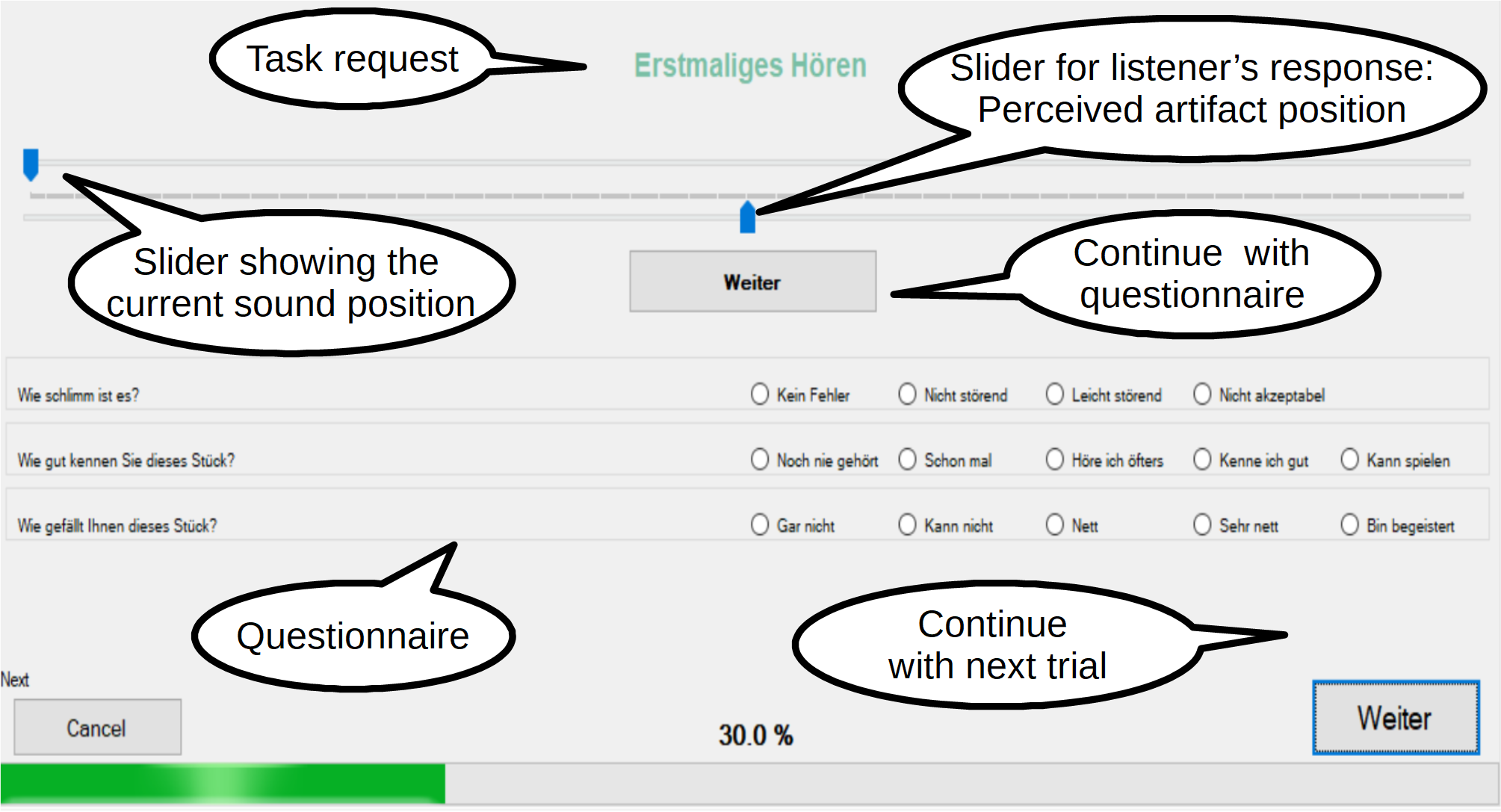} 
	\end{center}
	\caption{\change{The interface used in the experiment. See text for more details.
		\label{fig:Screenshot}}}
\end{figure}

\paragraph{Task}
In each trial, subject listened to a sound stimulus and was asked to pay attention to a potential
artifact \change{(see Fig.~\ref{fig:Screenshot})}. A slider scrolled horizontally while the sample was played indicating the current position within a stimulus. The subject was asked to tag the artifact's position by aligning a second slider with the begin of the perceived artifact. Then, while listening again to the same stimulus, the subject was asked to confirm (and re-align if required) the slider position and answer three questions: 
\begin{enumerate}
   \item \emph{Severity (S)}: How poor was it ("Wie schlimm ist es")? The possible answers were: (0) no issue ("Kein Fehler"), (1) not disturbing ("Nicht störend"), (2) mildly disturbing ("Leicht störend"), and (3) not acceptable ("Nicht akzeptabel").
   \item \emph{Familiarity (F)}: How familiar are you with this song ("Wie gut kennen Sie dieses Stück"): (0) never heard before ("Noch nie gehört"), (1) I have heard it before ("Schon mal gehört"), (2) I often listen to ("Höre ich öfters"), (3) I know it well ("Kenne ich gut"), and (4) I can play/sing it ("Kann ich spielen/singen").
   \item \emph{Liking (L)}: How do you like this song ("Wie gefällt Ihnen dieses Stück"): (0) not at all ("Gar nicht"), (1) I can not tell ("Kann nicht sagen"), (2) nice ("Nett"), (3) very nice ("Sehr nett"), and (4) amazing ("Bin begeistert").
\end{enumerate}
The questions were answered by tapping on the corresponding category. Then, the subject continued with the next trial by tapping the "next" button.

Before the experiment, the subject was informed about the purpose and procedure of the experiment and an exemplary reconstruction was presented. Any questions with respect to the procedure were clarified.

\paragraph{Conditions} 
Three conditions were tested. For the \textit{inpainting condition}, the song was corrupted at a random place with the gap of 1~s duration and then reconstructed with the default parameters from Tab.~\ref{tab:parameters}. The 
reconstructed song was cropped 2 to 4 seconds (randomly varying) before and 
after the gap resulting in samples of 5 to 9-s duration. \change{The gap was not allowed to be within the first and last 30 s of the song, but the inpainting was allowed to use the full song for processing.} For the \textit{reference 
condition}, the song was cropped at a random place with a duration varying from 5 
to 10 seconds. The reference condition did not contain any artifact and was used to estimate the sensitivity of a subject. For the \textit{click condition}, a click was superimposed to the song at 
a random position and the result was cropped 2.5 to 4.5 s before and after the 
click's position resulting in samples of 5 to 9-s duration. The artifact in this condition was used as a reference artifact and was clearly audible.\change{\footnote{\change{For other music genres like electronic music, the click might not be always audible and an other type of reference artifact would have been required.}}} 

In total, three inpainted, one reference, and one click conditions were created per song. 

\change{The combination of genres}, songs-per-genre, and conditions-per-song resulted in a 
block of 120 stimuli. All stimuli were normalized in the level (the click 
condition was normalized before superimposing the click). Within the block, the 
order of the stimuli and conditions was random.

Each subject was tested with two blocks, resulting in 240 trials per subject in 
total. Subjects were allowed to take a break at any time, with one planned break 
per block. For each subject, the test lasted approximately 2.5 hours. 

\subsection{Results}
\begin{figure}[htb!]
\begin{center}
\includegraphics[width=0.75\linewidth]{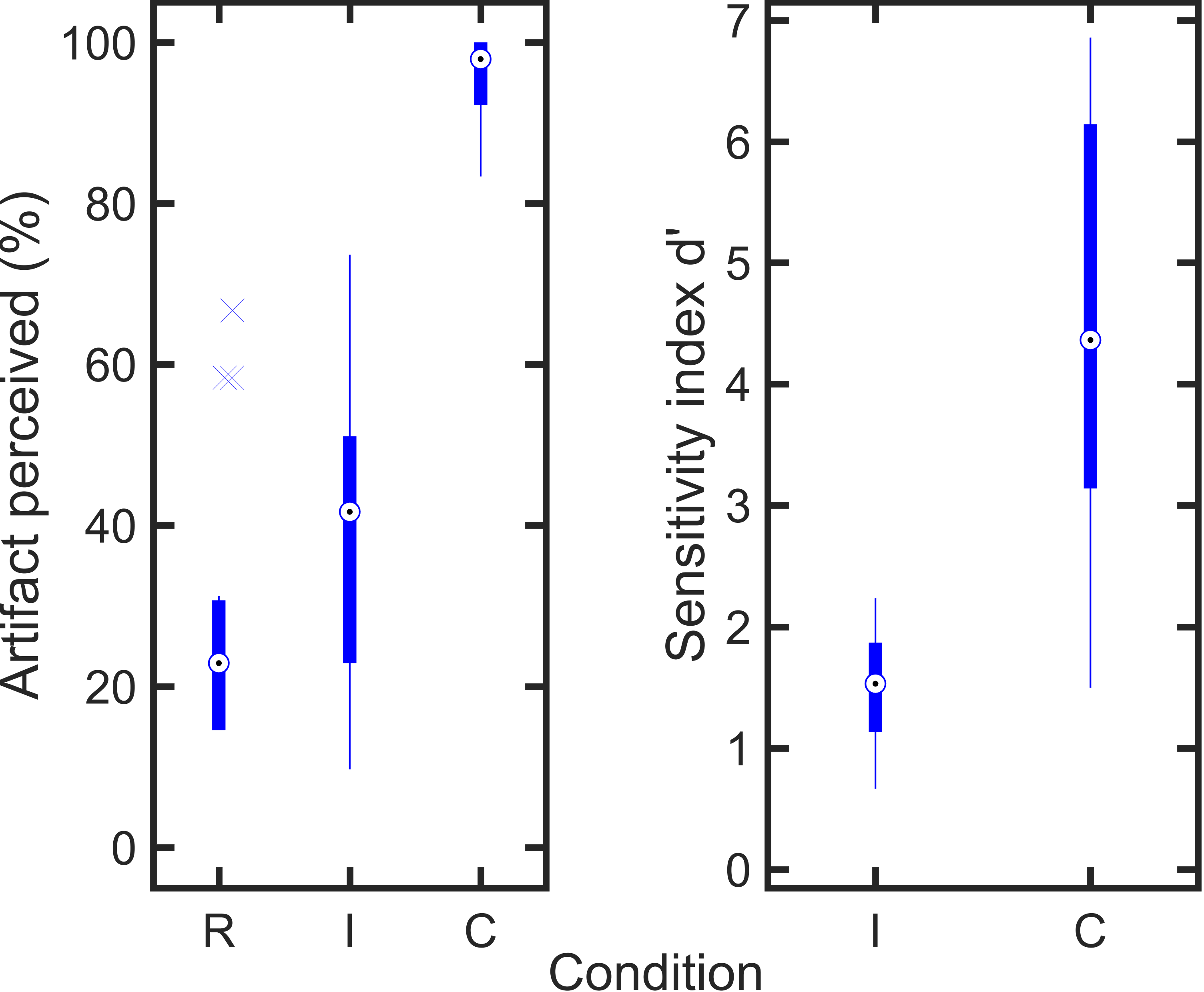} 
\end{center}
\caption{Detectability of artifacts is much lower than those of clicks but slightly higher than guessing. Left: Statistics of the rate of perceived artifacts across all subjects. Right: Statistics of the sensitivity index $d'$, i.e., the inpainting-detection rate relative to the false-alarm rate, across all subjects. $d'$ of 1 corresponds to the chance rate. Condition: Reference (R), inpainted (I), and click (C). Statistics: Median (circle), 25\% and 75\% quartiles (thick lines), coverage of 99.3\% (thin lines, assuming normal distribution), outliers (crosses, horizontally jittered for a better visibility). \label{fig:EvalPercentage}}
\end{figure}

\paragraph{Detection rate of the artifacts}
The detection results are shown in the left panel of Fig.~\ref{fig:EvalPercentage}. The average detection rates for the click, inpainting, and reference conditions were $95.6\pm5.0\%$, $40.1\pm19.2\%$, and $28.6\pm17.9\%$, respectively. The high detection rate and small variance in the click condition demonstrate a good attention of our subjects, for whom even a single click was clearly audible. The clearly non-zero rate in the reference condition shows that our subjects were highly motivated in finding artifacts. The detection rate in the inpainted condition was between those from the reference and click conditions. Note that the reference condition did not contain any artifacts, thus, the artifact's detection rate in that condition is here referred to as the false-alarm rate.

The large variance of the false-alarm rate shows that it is listener-specific. Thus, for further analysis, the detection rates from the inpainted condition were related to the listener-specific false-alarm rate, i.e., the sensitivity index $d'$ was used \cite{macmillan2004detection}. The false-alarm rate can be considered as a reference for guessing, thus, $d'=1$ indicates that the artifacts was detected at the level of chance rate. The right panel of Fig.~\ref{fig:EvalPercentage} shows the statistics of $d'$ for the inpainting and the click conditions. For the click condition, the average across all subjects was $4.36\pm1.91$, again demonstrating a good detectability of the clicks. For the inpainting condition, the average $d'$ was $1.49\pm0.42$, i.e., slightly above guessing ($d'= 1$). A t-test performed on listener's $d'$s showed a significant ($p = 0.0005$) difference from guessing, \change{indicating that the our listeners, as a group, were able to often detect the artifacts better than guessing. A 
listener-specific analysis, however, showed that only seven out of our 15 subjects were able to detect the inpainting better than chance, as revealed by a 2-by-2 contingency table analysis with the false-alarm and inpainting-detection rates evaluated at a significance level of 0.05. }

\paragraph{Influence of familiarity on the detectability}
A natural question that arises for this method is, in how far familiarity with a song will influence the detectability of the artifacts. While a comprehensive answer to this question is beyond the scope of this paper and would require a whole new study, here we aim at a brief impression for our subject pool. 

\begin{figure}[htb!]
\begin{center}
\includegraphics[width=0.75\linewidth]{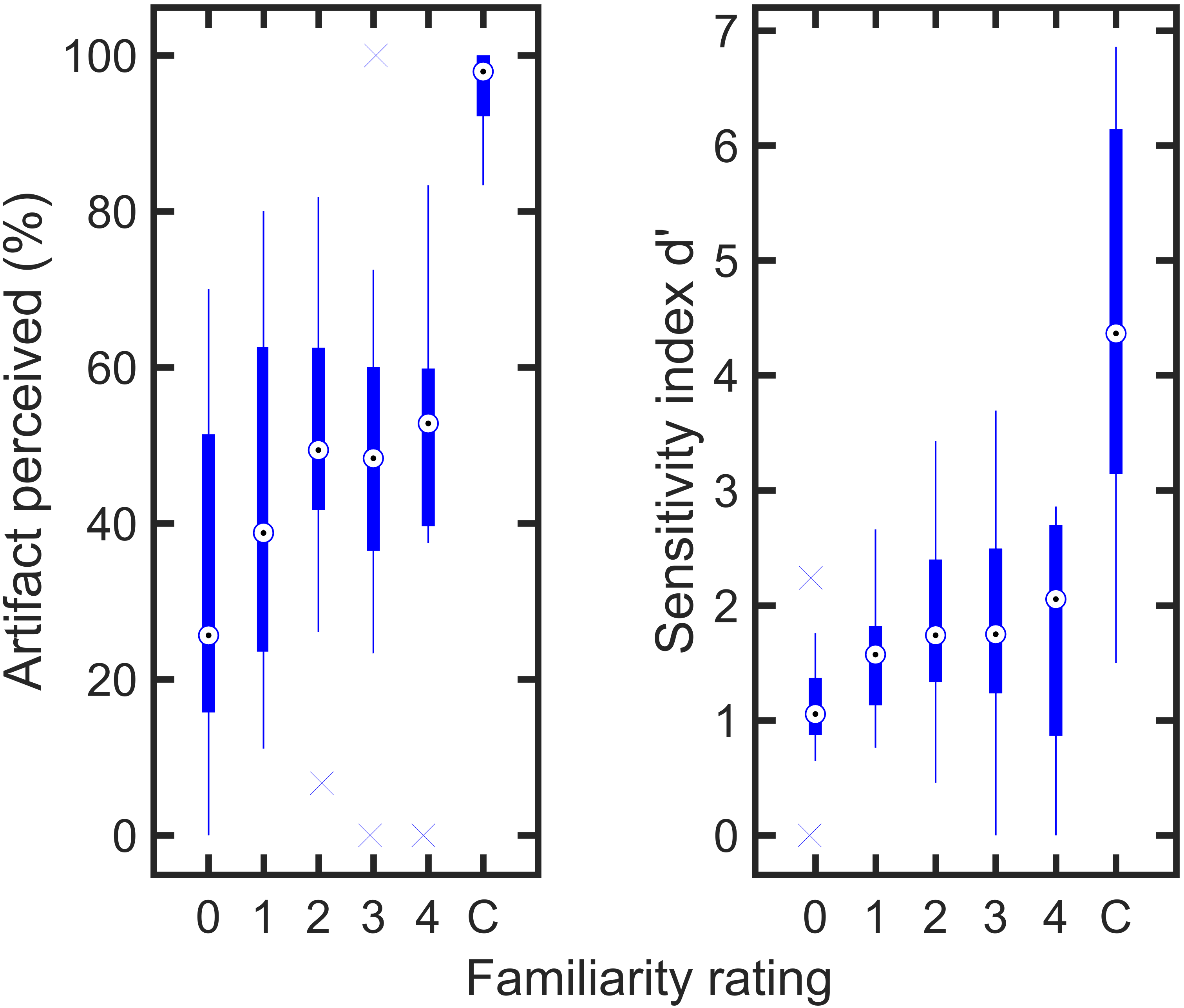} 
\end{center}
\caption{Detectability is not much related with the familiarity. Left: Statistics of the rate of perceived artifacts across all subjects as a function of the familiarity rating. Right: Statistics of the sensitivity index $d'$  as a function the familiarity rating. All other conventions as in Fig.~\ref{fig:EvalPercentage}.
\label{fig:EvalDetectFam}}
\end{figure}

Fig.~\ref{fig:EvalDetectFam} shows the detection rate (left panel) and the $d'$ (right panel) as functions of the familiarity ratings. 
While there seems to be a correlation of detectability and familiarity, surprisingly the link is not very strong. Arguably, there seems to be nearly no difference in the detection rates between songs rated with familiarity rating between 2 ("I often listen to") and of 4 ("I can sing/play it"), while there seems to be some difference to the other ratings of less familiarity. Interestingly, even for the very familiar songs the detection rate is much lower than for clicks and the $d'$ is only twice as large as that for the the chance rate.

\begin{figure}[htb!]
\begin{center}
\includegraphics[width=0.98\linewidth]{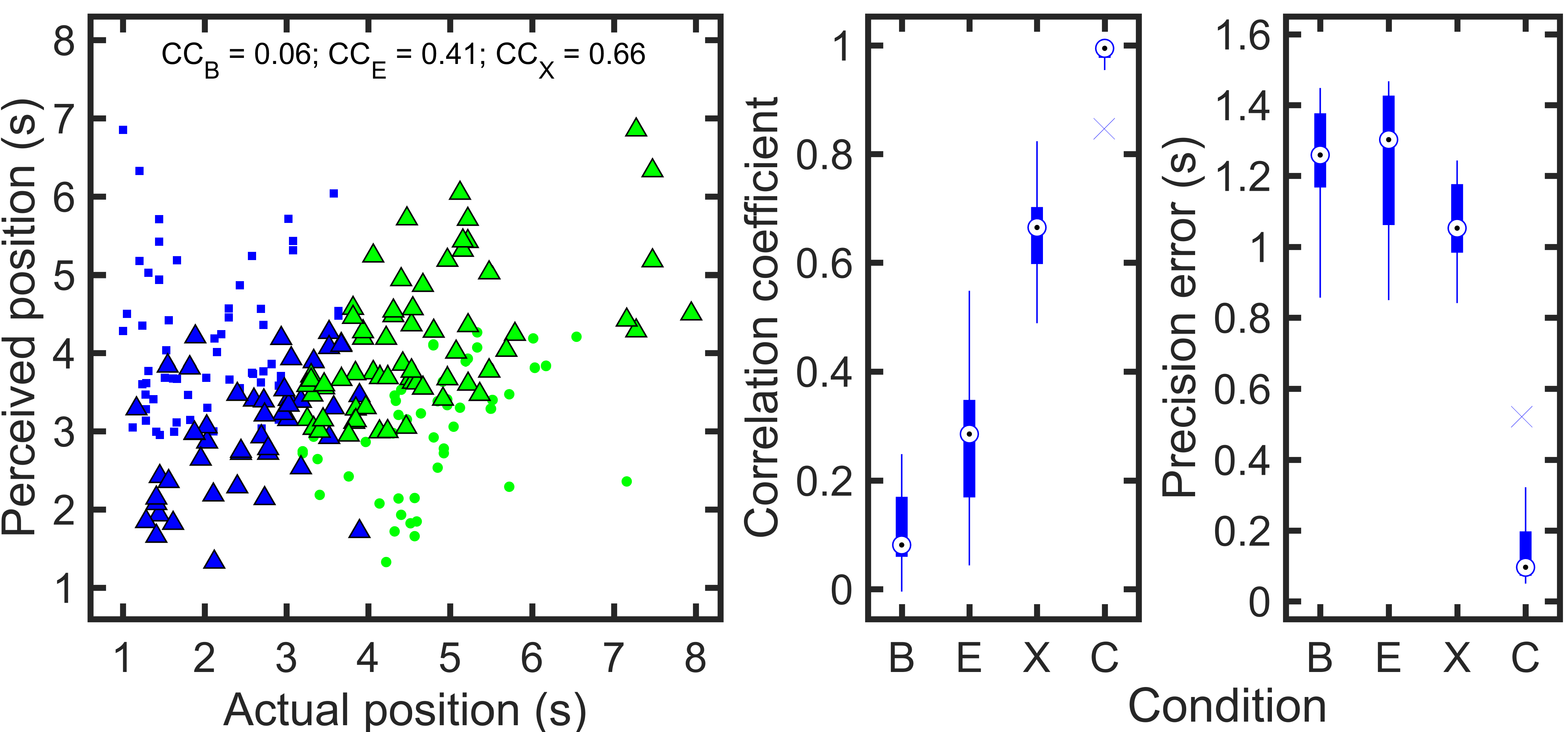} 
\end{center}
\caption{The position of perceived artifacts is weakly correlated with their actual position. Left: Perceived versus actual artifact's begin and positions (blue squares and green circles, respectively) for an exemplary subject. Triangles show the "best choice", i.e., perceived positions being nearer to either begin or end actual positions. Center: Statistics of the correlation coefficients for all subjects. Right: Statistics of the precision error for all subjects. B, E: perceived position versus begin and end of the artifact, respectively, in the inpainting condition. X: perceived position versus "best choice" in the inpainting condition. C: perceived position of the click in the click condition. $CC_B$, $CC_E$, $CC_X$: cross-correlation coefficient for the condition B, E, and X, respectively, of the exemplary listener. All other conventions as in Fig.~\ref{fig:EvalPercentage}). \label{fig:EvalCorr}}
\end{figure}

\paragraph{Detection of the artifact position}
\change{
	Subjects who successfully detected an artifact should be able to provide an information about its position within the stimulus, i.e., the perceived position of the artifact should correlate with its actual position. The left panel of Fig.~\ref{fig:EvalCorr} shows the perceived positions plotted versus the actual positions of the artifacts, for an exemplary average listener. The reported perceived artifact's position might refer to gap's begin or end, with the choice even varying from stimulus to stimulus. Thus, we correlated the reported position with the begin, the end, and the nearer of the two positions (referred to as "best choice"). The "best choice" positions are highlighted by triangles. 
	
	Across all subjects, correlation coefficients' statistics is shown in the center panel of Fig.~\ref{fig:EvalCorr}. The moderate correlations indicate that as soon as our subjects detected an artifact, they had some estimate of its position within the stimulus. In contrast, for the clicks, the high correlation indicates that our subjects were able to exactly determine and report the position of the click artifact.

	In order to determine the precision in the reporting the artifact's position, we also calculated the difference between the perceived and actual artifact's position. The standard deviation of these differences calculated for a subject is referred to as the precision error. Their statistics across subjects is shown in the right panel of Fig.~\ref{fig:EvalCorr}. For the click condition, the average precision error across all subjects was $157\pm130$\,ms. It describes the procedural precision of subjects within our task. For the inpainting condition, the average precision error considering the artifact's begin, end, and "best choice" as the actual position was $1232\pm180$\,ms, $1247\pm199$\,ms, and $1069\pm115$\,ms, respectively. The "best choice" shows the lowest precision errors, being more than six times larger than the procedural precision error. This indicates that even if detected, our subjects had large difficulties to determine the artifact's position and these difficulties did not originate from the 
task. 
}

\begin{figure}[htb!]
\begin{center}
\includegraphics[width=0.95\linewidth]{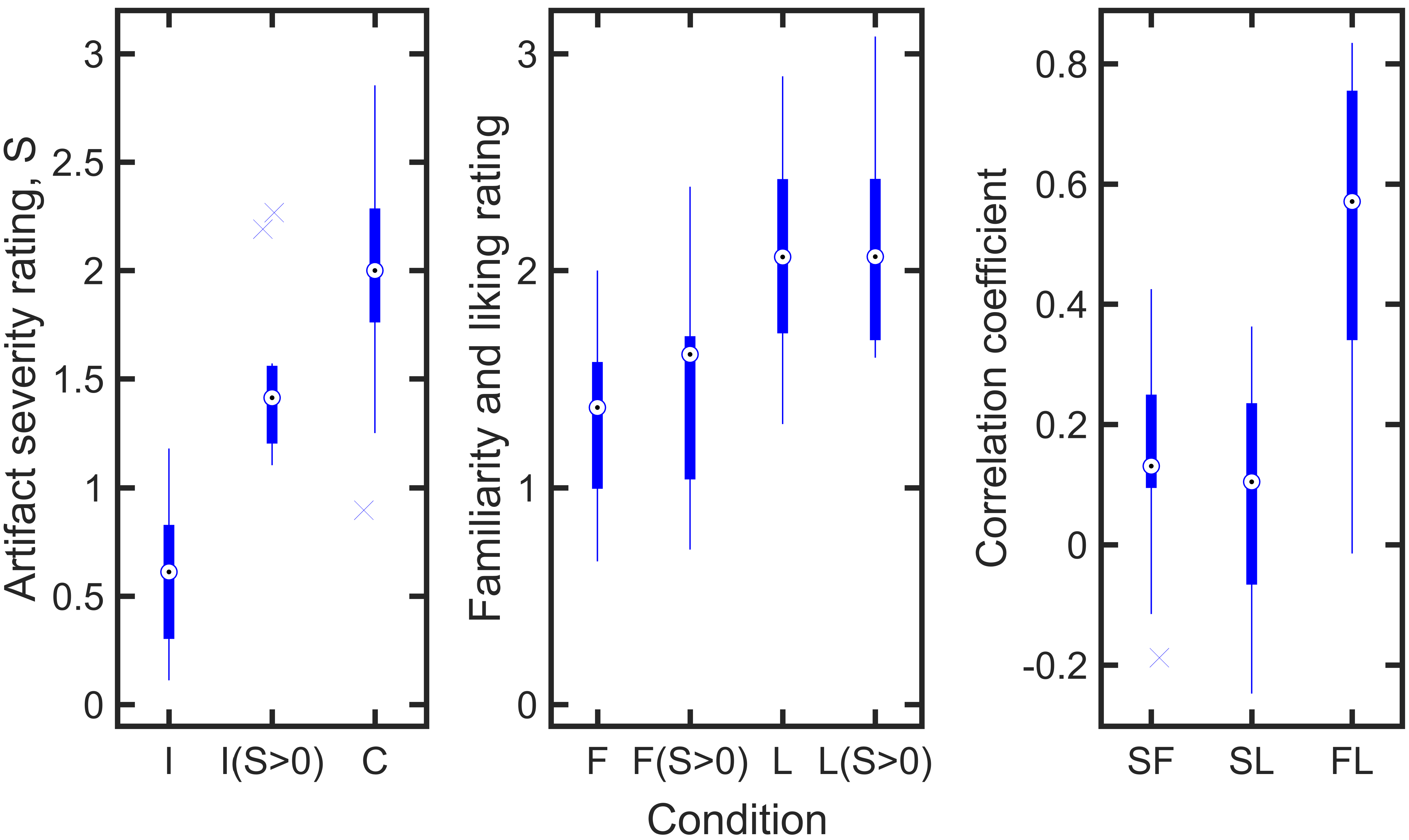} 
\end{center}
\caption{Statistics of ratings across all subjects. Left: severity ratings (S). Center: Familiarity (F) and liking (L) ratings. Condition: Inpainted (I), click (C), ratings considering perceived artifacts only (S>0). Right: Statistics of Pearson's correlation coefficients between S and F (SF), S and L (SL), as well as F and L (FL). All other conventions as in Fig.~\ref{fig:EvalPercentage}. \label{fig:Ratings}}
\end{figure}

\paragraph{Disturbance rate of detected artifacts}
Finally, we have analyzed the ratings we have collected. The left panel of Fig.~\ref{fig:Ratings} shows the statistics of the severity ratings reported in the inpainted and click conditions. For the click condition, most of the ratings were between 1 ("not disturbing") and 3 ("not acceptable") with an average across all subjects of $2.00\pm0.55$. This indicates that on average, our subjects rated the clicks as disturbing. In contrast, for the inpainted condition, most of the ratings were between 0 ("no issue") and 1 ("not disturbing") with an average of $0.60\pm0.33$. This indicates that on average, our subjects rated the inpainting results halfway between "no issue" and "not disturbing".

This analysis considered all inpainted stimuli so far, ignoring the fact that for some of them our subjects detected the artifact and for some not. A statistic of severity ratings considering detected artifacts only (i.e., $S>0$) is shown in the center part of the left panel in Fig.~\ref{fig:Ratings}. The average across all subjects was $1.46\pm0.35$. This is higher than the average considering all severity ratings, but still significantly ($p=0.0002$) lower than the severity of the clicks as revealed by a paired t-test calculated between the ratings for clicks and inpainted but detected artifacts. This indicates that even when the inpainting artifacts were perceived, their severity was rated significantly lower than that of the clicks.

\paragraph{Influence of the familiarity on the severity}
The stimulus' familiarity and liking might also have influenced our experimental outcome. The average ratings for the familiarity and liking are shown in the center panel of Fig.~\ref{fig:Ratings}. Most of the familiarity ratings were between category 1 ("I have heard it several times") and 2 ("I often listen to"), with an across-subject average of $1.23\pm0.41$. Considering the perceived artifacts only  (i.e., $S>0$), the average increased to $1.44\pm0.45$. This increase was significant ($p=0.022$, paired t-test on all and the perceived only ratings), indicating that our subjects were slightly more familiar with stimuli containing detectable artifacts. The liking ratings were mostly between 1 ("I cannot tell") and ("very nice"), with an average of $2.10\pm0.43$. Considering perceived artifacts only, the average increased to $2.13\pm0.44$. This increase was not significant ($p=0.68$, paired t-test between all and the perceived only ratings). As it seems, the artifact's detectability was not related to the 
song liking.

The link between the severity of an artifact and the familiarization and/or liking ratings was further investigated by calculating the Pearson's correlation coefficients between the severity and other ratings. The right panel of Fig.~\ref{fig:Ratings} shows the group statistics of the correlation coefficients, which, on average, were $0.14\pm0.16$ and $0.09\pm0.17$ for the correlation of severity with familiarity and liking, respectively. Such low correlations indicate that neither the familiarity nor liking was clearly linked with the perceived artifact's severity. Out of curiosity, also the correlation between the familiarity and liking was calculated, resulting in an across-subject average of $0.54\pm0.26$. This correlation indicates a good link between the familiarity and liking of our stimuli, but also raises evidence that familiarity and liking are not fully equivalent. 

\section{Conclusions}

We have introduced a method for the restoration of audio signals in the presence
of corruption/loss of data over an extended, connected period of time. Since,
for complex audio signals, the length of the lost segment usually prohibits
the inference of the correct data purely from the adjacent reliable data, our
solution is based on the larger scale structure of the underlying audio
signal. The reliable data is analyzed, detecting spectro-temporal
similarities, resulting in a graph representation of the signal's temporal
evolution that indicates strong similarities. Inpainting of the lost data is
then achieved by determining two suitable transitions between the border
regions around the  corrupted signal segment and a region that is considered
to be similar. In other words, the algorithm jumps from shortly before the
\emph{gap} to a similar section of the audio signal and, after some time,
back to a position shortly after the gap, effectively exchanging the
corrupted piece with a suitable substitute. Consequently, the  algorithm is
capable of efficiently exploiting naturally occurring redundancies  in the
reliable data.

In order to test the efficiency of our algorithm, we have conducted a psychoacoustic evaluation. The results show that our listeners were able to detect $40\%$ of the artifacts implying that our method completely fooled our listeners more than $60\%$ of the time. Our listeners showed a false-alarm rate of $28\%$, indicating that sensitivity of correctly detecting a gap was with $d'=1.49$ rather low (as compared with $d'=4.36$ for well-detectable clicks and with $d'=1$ for the chance rate). \change{In fact, listener-specific analysis showed that only seven out of 15 tested listeners were able to detect the inpainting on a statistical significant level. }  Our study showed two additional quality signs of our method. First, the detected artifacts were rated on average between ``not disturbing'' and ``mildly disturbing''. Second, even though detected, our subjects \change{only vaguely determined the artifact's position, with the six-fold detection precision error than that in the reference condition. While our 
test was limited to four music genres, they covered many music structures usually found in other genres. However, inpainting performed on a very different genre like the contemporary electronic music might have led to different results, both numeric and perceptual.}

Besides having built and tested a novel audio inpainting algorithm, it is
worth noting that the graph constructed with our method gives an intuitive
analysis of the signal at hand, exposing self-similarities and global
structure and can be used for a number of different purposes. For example, a
song can be  re-composed by following the edges of the graph while respecting
the global music structure. \change{Multiple matches in a highly repetitive song can be used as a tool for further song modifications, offering a creative tool for algorithmic composing, e.g., in the field of contemporary electronic music. } 

\change{Similarity graphs can be used in many
applications, thus, it is important to further improve this kind of signal representation.}
Hence, future work includes closing the gap between the internal similarity
measures and human hearing by incorporating perceptually motivated similarity
measures derived, possibly, from a perceptually-motivated
representation~\cite{DBLP:journals/corr/NecciariHBP16} or a computational
model of the auditory system~\cite{Irino:2006b}. Such a modification will
greatly improve the reliability of the algorithm and its results. It seems
worth noting, however, that even after considering an auditory model,
reliable retrieval of strongly context-sensitive data such as speech and
singing voice will require additional contextual information and might be
better achieved by a generative approach~\cite{ saino2006hmm}, applied after
separating voice and music in the signal~\cite{li2007separation}.



\section*{Acknowledgment}
\change{We thank the reviewers and the editor for their review this publication and their helpful suggestions.} We thank Pierre Vandergheynst for his support during this project. His ideas and 
suggestions have helped significantly to this contribution.

This work has been supported by the Swiss Data Science Center and by the Austrian Science Fund (FWF) projects FLAME 
(\textit{Frames and Linear Operators for Acoustical Modeling and Parameter 
Estimation}; Y 551-N13) and MERLIN (\textit{Modern methods for the restoration of 
lost information in digital signals}; I 3067-N30)

\begin{table*}[ht!]
\begin{center}
\begin{tabular}{|l|l|l|l|}
   \hline
  \bf{ Quantity} &  \bf{Variable used }&  \bf{Default value} &  \bf{Unit }\\
   \hline
   \hline
   \multicolumn{4}{|c|}{\emph{Audio features}} \\
   \hline
   Maximum sampling frequency & $\xi_{s,\text{max}}$ & $12'000$ &$Hz$ \\
   \hline
   Size of the patch  & $a$ & $128$ & samples \\
   \hline
   Number of frequencies & $M$ & $1024$ & - \\
   \hline
   Length of the window & $L_w$ & $M$ & samples \\
   \hline
   Type of window & - & 'Itersine' & - \\
   \hline
   Dynamic range & $p$ & $50$ & dB \\   
   \hline
   Trade-off between the amplitude and phase & $\lambda$ & $3/2$ & - \\
   \hline
   \multicolumn{4}{|c|}{\emph{Graph}} \\   
   \hline
   Initial number of neighbors & $K$ & $40$ & - \\
   \hline
   Kernel length & $L_k$ & $40$ & - \\   
   \hline   
   Hard threshold for the weight matrix & $t_w$ & $2$ & - \\   
   \hline
   \multicolumn{4}{|c|}{\emph{Optimization}} \\   
   \hline   
   Regularization parameter 1 & $\gamma_1$ & $1$ & - \\   
   \hline   
   Regularization parameter 2 & $\gamma_2$ & $100$ & - \\   \hline
\end{tabular}
 \end{center}
   \caption{\label{tab:parameters}Default parameters of the algorithm}
\end{table*}

\bibliographystyle{IEEEtran}
\bibliography{biblio}

\end{document}